\documentclass[sigconf, nonacm]{acmart}

\usepackage[binary-units]{siunitx}
\DeclareSIUnit{\x}{\times}
\setcopyright{none}
\renewcommand\footnotetextcopyrightpermission[1]{}

\usepackage{booktabs} 
\usepackage{algorithm}
\usepackage{algpseudocode}
\usepackage{mdframed}
\usepackage{array}
\usepackage{tabularx}
\usepackage{nicefrac}
\usepackage{enumerate}
\usepackage{enumitem}
\usepackage{caption}
\usepackage{subcaption}
\usepackage{lipsum}
\usepackage{float}
\usepackage{comment}
\usepackage{xcolor}
\usepackage{amsmath}
\usepackage{amsthm}
\usepackage{pifont}
\usepackage{hyperref}
\usepackage{xspace}
\usepackage{graphics}
\usepackage{graphicx}
\usepackage{multirow}
\usepackage{outlines}
\usepackage{xurl}
\usepackage{stfloats}

\newcommand{\ie}{{\it i.e.,}\xspace}
\newcommand{\eg}{{\it e.g.,}\xspace}

\newcommand{\BigO}{\mathcal{O}}

\newcolumntype{L}[1]{>{\raggedright\arraybackslash}p{#1}}
\newcolumntype{C}[1]{>{\centering\arraybackslash}p{#1}}
\newcolumntype{R}[1]{>{\raggedleft\arraybackslash}p{#1}}

\newcommand{\parab}[1]{\vspace{0.06in}\noindent{\bf #1}}

\newcommand{\greybox}[1]{\vspace{0.1in}\begin{mdframed}[backgroundcolor=gray!15]{\vspace{0.05in}#1\vspace{0.05in}}\end{mdframed}}

\newcommand{\vartext}[1]{\text{\textit{#1}}}

\begin{document}

\title{Resource Allocation in Serverless Query Processing}

\author{Simon Kassing, Ingo M\"uller, Gustavo Alonso}
\affiliation{%
  \institution{Systems Group, Department of Computer Science, ETH Z\"urich \\ Z\"urich, Switzerland}
}
\email{{simon.kassing,ingo.mueller,alonso}@inf.ethz.ch}

\begin{abstract}
Data lakes hold a growing amount of cold data that is infrequently accessed, yet require interactive response times. Serverless functions are seen as a way to address this use case since they offer an appealing alternative to maintaining (and paying for) a fixed infrastructure. Recent research has analyzed the potential of serverless for data processing. In this paper, we expand on such work by looking into the question of serverless resource allocation to data processing tasks (number and size of the functions). We formulate a general model to roughly estimate completion time and financial cost, which we apply to augment an existing serverless data processing system with an advisory tool that automatically identifies configurations striking a good balance---which we define as being close to the "knee" of their Pareto frontier. The model takes into account key aspects of serverless: start-up, computation, network transfers, and overhead as a function of the input sizes and intermediate result exchanges. Using (micro)benchmarks and parts of TPC-H, we show that this advisor is capable of pinpointing configurations desirable to the user. Moreover, we identify and discuss several aspects of data processing on serverless affecting efficiency. By using an automated tool to configure the resources, the barrier to using serverless for data processing is lowered and the narrow window where it is cost effective can be expanded by using a more optimal allocation instead of having to over-provision the design.
\end{abstract}

\pagestyle{plain}
\maketitle

\section{Introduction}

Serverless computing, now a feature offered by all major cloud providers~\cite{aws-lambda, google-cloud-functions, azure-functions}, has been touted as a major step forward in terms of more efficient computing \cite{Serverless-CACM19,Serverless-CACM21}. In spite of the limitations of serverless platforms \cite{OpenLambda16, Peeking-ATC18, Hellerstein-CIDR19}, recent results indicate that analytical query processing over serverless is feasible but that it has a very narrow window where it remains cost effective. The work on Starling~\cite{starling} and Lambada~\cite{lambada} show that, in current commercial offerings, the maximum throughput is a few tens of queries per hour. Beyond that, it is better to use a fixed infrastructure as it costs less per computation time unit. Nevertheless, these systems show that query processing over serverless can be competitive for interactive, occasional use. For instance, it can be an useful extension to data lakes such as Databrick's Lakehouse~\cite{databricks}, Microsoft Azure's Data Lake~\cite{azure-data-lake}, or Amazon's Redshift~\cite{amazon-redshift}. Over large swathes of the data in these large repositories, analytical queries are run only infrequently resulting into what is referred to as "cold" data. Loading such data into traditional databases for processing is costly and impractical. Using a data lake infrastructure is more economical but still incurs too high costs for only occasional use. This is where serverless query processing can be very useful. Compared with using virtual machines, serverless functions have much lower latency to start, are more scalable, and are charged at a smaller time granularity (e.g., 1~ms~\cite{aws-lambda-pricing}).

Provisioning resources in the cloud for query processing is not easy \cite{leis2021towardscostoptimal}. For serverless, the narrow effective window and the limited choices given to the user make the task even more difficult \cite{Peeking-ATC18}. However, for occasional use over the cold data, such provisioning is critical to keep the system cost-effective. The two obvious extremes yield little practical value to the user: the cheapest choice (small number of the smallest functions) finishes too slow; the fastest (a large number of the largest functions) is too expensive. Our goal in this paper is to develop an advisory tool that can recommend suitable configurations for serverless query processing striking a good balance between completion time and financial cost. 

To do this, we propose a model similar to \cite{leis2021towardscostoptimal}'s approach for VMs, that considers the major factors affecting completion time and cost in serverless query data processing. It includes processing and network overheads (as \cite{leis2021towardscostoptimal}, except using data and rate rather than CPU hours for processing), and as well accounts for startup time, exchange overhead, and request cost. The latter three factors are relevant for serverless data processing due to (a) the large number of workers involved, (b) the brevity of interactive queries, and (c) communication through storage.

We use this model to augment an existing query data processing system, Lambada~\cite{lambada}, with an advisory tool to provision serverless functions: given a query workload, the tool approximates the number of functions and their size to achieve a good balance between query completion time and the cost incurred. It does so by choosing configurations close to the "knee" of the Pareto frontier of these two dimensions. We show that obtaining accurate predictions of the run time or cost on serverless is very difficult due to a variety of system aspects that cannot be controlled by the user \cite{Peeking-ATC18,OpenLambda16}. However, we also show that it is still possible to recommend a reasonable configuration, thereby significantly simplifying the job of the user when using serverless for query processing over data lakes and expanding the narrow cost effectiveness of serverless query processing. 


\section{Why serverless data processing?}
\label{sec:why}

Serverless has been proposed as an alternative to using virtual machines to process interactive queries on cold data~\cite{lambada, starling}. \textit{Cold} data is so infrequently accessed that it is not cost-effective or practical to keep infrastructure on hot stand-by for processing it. Yet, users still expect \textit{interactive} response times which requires sufficient computing power. For other query processing tasks, serverless might be too expensive. There are two reasons for this:

\parab{Billing method.} Virtual machine instances are billed both at coarser time granularity and with a minimum time across all cloud providers. Both AWS~\cite{aws-ec2-pricing} and Google Cloud~\cite{gc-instance-pricing} charge at a one second granularity, with a minimum of 60 seconds. Azure bills at either second~\cite{azure-container-instance-pricing}, or minute~\cite{azure-vm-instance-pricing} granularity---we could not find whether a minimum applies. Both factors substantially increase the cost of interactive queries whose completion time is in seconds. Suppose a query takes \SI{6.5}{\second} to complete. This would be billed \SI{8}{\percent} more expensive at one-second granularity, and \SI{823}{\percent} more expensive at a one-minute granularity. In contrast, serverless functions are billed at very fine time granularity (AWS/Azure: \SI{1}{\milli\second}~\cite{aws-lambda-pricing, azure-functions-pricing}, Google Cloud: \SI{100}{\milli\second}~\cite{gcf-pricing}) and with a small minimum billing duration per invocation (AWS: \SI{0}{\milli\second}~\cite{aws-lambda-pricing}, Google Cloud/Azure: \SI{100}{\milli\second}~\cite{gcf-pricing,azure-functions-pricing}).

\parab{Startup delay.} Due to data coldness, when an interactive query comes in, workers must be started as soon as possible in order to scale to process the data in a timely fashion. VM instances take significantly longer to start up in comparison to serverless functions, in the range of several tens of seconds to start up~\cite{vm-startup-times-analysis}. This is unacceptable for an interactive query that takes seconds to run. In contrast, serverless functions can start within a few hundred milliseconds~\cite{catalyzer-serverless-startup} and a lot of effort is being invested in shortening that even further \cite{catalyzer-serverless-startup,SAND18,FaaSNet21}.


\section{Provisioning and configuration}
\label{sec:provisioning}

\begin{table}[t]
    \centering
    \renewcommand{\arraystretch}{1.3}
    \newcommand{\GBrange}[2]{\SIrange[range-phrase=--,range-units=single]{#1}{#2}{\gibi\byte}}
    \begin{tabular}{| p{1.65cm} | C{1.85cm} | C{1.85cm} | C{1.85cm} | } \hline
         & \textbf{AWS} & \textbf{GCP} & \textbf{Azure} \\ \hline \hline
        \textbf{Memory} & Configurable \GBrange{0.125}{10} \cite{aws-lambda-pricing} & Configurable \GBrange{0.125}{8} \cite{gcf-pricing} & Configurable \GBrange{1.5}{14} \cite{azure-functions-pricing} \\ \hline
        \textbf{Compute} & Based on memory (1769 MiB/vCPU) \cite{aws-lambda-config} & Based on memory \cite{gcf-pricing} & Based on memory~\cite{azure-functions-pricing} \\ \hline
        \textbf{Network bandwidth} & \multicolumn{3}{c|}{No information available~\cite{aws-lambda-pricing, gcf-pricing, azure-functions-pricing}} \\ \hline
        \textbf{Local\newline disk size\newline (scratch)} & Constant (\SI{512}{\mebi\byte}) \cite{aws-lambda-faq} & Stored within the memory itself~\cite{gcf-file-system} & Constant (\SI{500}{\mebi\byte})~\cite{azure-kudu-temp-storage} \\ \hline
    \end{tabular}
    \caption{\em Current serverless configuration offerings.}
    \label{tab:cloud-offering}
    \renewcommand{\arraystretch}{1}
    \vspace{-15pt}
\end{table}

A user must decide what type of serverless workers to start and how many, in the same way as one would for a VM-based data processing system \cite{Microsoft-Protean20}. They differ however on the range of available configurations. For traditional virtual machine instances, cloud providers offer users the choice among a wide range of machine configurations, varying four key parameters \cite{leis2021towardscostoptimal}: number of cores (\textit{compute}), network bandwidth (\textit{network}), DRAM (\textit{memory}), and SSD (\textit{local disk size}). 
In contrast, serverless configurations are far more limited. Users can only select the memory size, from which some of the other properties are derived (Tab.~\ref{tab:cloud-offering}). 

\greybox{Based on the current cloud offering, in this work we define the \textbf{provisioning of serverless data processing systems} as the choice of the \textbf{number of serverless workers} ($W$) and their \textbf{memory size in mebibyte} ($M$).}


\section{Estimation challenges}
\label{sec:prediction-challenges}

To recommend a suitable configuration, we build a model (\S\ref{sec:general-model}) encompassing key characteristics of a serverless query execution system. The model does not try to make an accurate prediction of running time or cost but to differentiate between configurations. Estimating running time and cost in serverless is actually quite difficult (\S\ref{sec:into-practice}, \S\ref{sec:large-benchmark}) due to the many uncertainties associated with serverless execution \cite{Microsoft-study20,Peeking-ATC18}. In this section, we highlight both the fundamental prediction limitations and practical challenges inherent to serverless data processing to define the problem space.


\subsection{Data-processing-related limitations}
\label{sec:data-processing-limits}

At the core of query optimization is the need to predict the cost of an operation, for which often metadata and statistics on the data (selectivity, distribution, etc.) are used. It is an open question how to do this over the vast amounts of data stored in a data lake. There have been considerable efforts, using zone maps~\cite{Snowflake} or gradual transformations of the data~\cite{MIT-CIDR22}, as well as table formats to include cost-based optimization metrics~\cite{iceberg}. Lack of such data processing knowledge results in the first source of potential inaccuracies when trying to guess the running time of a query. In our case, these are the two factors that introduce the largest distortions:

\begin{itemize}[leftmargin=10pt,itemsep=2pt,topsep=2pt]

    \item \textbf{Query selectivity.} The selectivity of an operator determines the amount of data the following operators process. As a result, the cost of intermediate operators in the query plan is difficult to gauge and can only be done on average or as a worst case.
    
    \item \textbf{Data distribution.} The performance of an operator (\ie process rate) is affected by the data distribution of the input. For instance, the population of internal data structures can result in longer processing time in aggregation operators relying on hash tables. If many input tuple keys are unique, the hash table will become full, causing collisions, resizing overhead, no longer fitting in caches, etc. This makes it difficult to estimate the cost of such operators even when they are on the leaves of the query plan. Moreover, the size of each individual tuple can vary; \eg \SI{100}{\mebi\byte} of \SI{10}{\byte} tuples compared to \SI{100}{\mebi\byte} of \SI{100}{\byte} tuples would result in 10$\times$ more operations.
    
\end{itemize}

We can account for these variances in two ways: (a) bounding the data size and effect on complexity, or (b) with additional knowledge about the data. In this project we opt for the former to avoid relying on functionality that is typically not available in data lakes. Note that adding such information (e.g., as suggested by \cite{MIT-CIDR22}) to the model will only make it more accurate. 


\subsection{Serverless platforms related limitations}

Serverless query processing systems make use of the many services a cloud provider offers, in particular, storage and serverless functions. The experienced quality of service, however, can vary: (a) the way the implementation and resource allocation of cloud services are opaque to their users, and (b) the resources behind these services typically are not reserved: they are shared among users and, thus, service quality can vary. The following are side effects observed in Amazon AWS influencing run time and cost:

\begin{itemize}[leftmargin=10pt,itemsep=2pt,topsep=2pt]

    \item \textbf{Start-up time variance (\textit{starting stragglers}).} The time from invocation until the worker is ready to start the computation is sometimes longer than usual, already creating \textit{starting stragglers}. We hypothesize this can be an effect of the caching strategy, as well as temporarily contention due to high demand of the shared infrastructure.
    
    \item \textbf{Storage read/write variance (\textit{storage stragglers}).} Serverless query processing systems make use of long-term storage (\eg S3~\cite{amazon-s3}) to read input and potentially write output, as well as to exchange data in-between computational stages. Infrequently, the read or write requests to S3 take considerably longer than expected (by as much as one order of magnitude). This can be due to a variety of reasons: \eg S3 replication, S3 caching, network packets being lost, or high demand (on either the function host or S3). Even if it occurs only 1\% of the time, if there are 100 workers, these tail failure probabilities become significant~\cite{starling, tail-at-scale} and create large variations on the running time.
    
    \item \textbf{Queueing service send/receive variance.} Queuing services (\eg SQS) can be used to exchange small messages between workers, or to the driver, for example if the query result is small enough. Infrequently, but especially with many simultaneous finishes, the workers will simultaneously send their result message to the SQS queue. This can lead to significant delay if the underlying application/transport stack starts timing out. This is a known problem~\cite{alvarez2020specializing} in parallel processing when multiple parallel workers finish at the same time and all try to communicate back with a centralized server.  
    
    \item \textbf{Computational contention.} A worker is allocated computational power proportional to its memory allocation. With limited computational power, in particular, less than one vCPU, even a single-threaded system combined with background computation can exhibit unpredictable computation patterns that can affect the observed network bandwidth and, thus, the time needed to read/write data.
    
    \item \textbf{Non-linear network performance.} In line with previous work~\cite{boxer}, we observe that lambda functions in AWS can exhibit an increased network rate for a very short period of time at the beginning of their execution. Whether this behavior is seen or not depends on whether the storage (S3) can supply data at this increased rate and whether the function has sufficient parallelism available to deal with the higher incoming data rate (both in terms of vCPU and the need to download multiples files). In practice, it greatly affects the ability to estimate how long a query will take. 
    
    \item \textbf{Polling requests.} When awaiting for a file from another worker to become available, a worker will poll to check when it is available. The more frequently the worker polls, the faster it is made aware and can start reading; however, this affects the price since, the more read requests, the higher the cost of the computation. The duration of waiting (and thus polling) is influenced by stragglers.
    
\end{itemize}

Because of these intrinsic difficulties, in this paper we do not aim to build a query cost predictor. Our model does accurately estimate running time and cost in many situations but not always. However, the estimations are good enough to be able to differentiate between configurations in terms of picking suitable ones in terms of balancing running time and cost. This is the basis for the advisory tool we propose.


\section{General estimation model}
\label{sec:general-model}

Rough estimations of cost/execution time on virtual machines have been recently proposed~\cite{leis2021towardscostoptimal}. However, such an approach does not directly translate to serverless functions due to: (a) the difference in the number of workers involved (10s for virtual machines, 100s for serverless functions), (b) the overhead incurred for data exchanges (typically through storage), and (c) the difference in task duration. Serverless operates in the few to tens of seconds range, whereas virtual machines in minutes to hours -- factors insignificant at a one hour scale become so at a one second scale \cite{Peeking-ATC18,Microsoft-study20}. In this section, we formulate an analytical model which incorporates the most important factors in our understanding that effect the completion time and financial cost of query processing using serverless functions. The model includes processing and network overhead similar to \cite{leis2021towardscostoptimal} (using data and rate instead of CPU hours for processing), but also incorporates the factors unique to serverless query processing systems, namely startup time, exchange overhead and request cost. The model we present is directly applicable to Lambada~\cite{lambada}, and we posit models in a similar vein can be applied to other systems to roughly estimate performance albeit with changes, for instance to account for different number of workers per stage (\eg for \cite{starling, flint}).


\subsection{Start-up}
\label{sec:model-start-up}

The largest serverless function is equivalent to at most a medium-tiered virtual machine instance (\SI{10}{\gibi\byte} of memory with 5.8 vCPUs for AWS). Thus, many serverless functions must be spawned to be able to process large amounts of data. A single driver to start the workers can only start workers at a limited rate. To speed up the start up phase, it has been proposed to use a two-level broadcast tree where the first workers start other workers ~\cite{lambada}. We model the starting of serverless workers by an invocation rate $R_{\vartext{1-inv}}$ and $R_{\vartext{2-inv}}$ (\ie how many workers can be started per second) for the main driver and a worker respectively. We define the invocation delay as $T_{\vartext{inv. delay}}$ (\ie how long from invocation till the worker is started). For each worker $i \in 1.. W$, we estimate its startup and ready time.

\smallskip
\noindent \textbf{One-level invocation:}
$$T_{\vartext{startup}}(i) = \frac{i}{R_{\vartext{1-inv}}} + T_{\vartext{inv. delay}}$$
$$T_{\vartext{ready}}(i) = T_{\vartext{startup}}(i)$$

\noindent \textbf{Two-level invocation:}
$$
T_{\vartext{startup}}(i) =\begin{cases}
\frac{i}{R_{\vartext{1-inv}}} + T_{\vartext{inv. delay}} & i \leq \sqrt{W} \\
\frac{g}{R_{\vartext{1-inv}}} + T_{\vartext{inv. delay}} + \frac{h}{R_{\vartext{2-inv}}} + T_{\vartext{inv. delay}} & \vartext{else}\end{cases}
$$

$$
T_{\vartext{ready}}(i) =\begin{cases}
T_{\vartext{startup}}(i) + \frac{\sqrt{W}}{R_{\vartext{2-inv}}} & i \leq \sqrt{W} \\
T_{\vartext{startup}}(i)  & \vartext{else}\end{cases}
$$

$g$ is the index of the worker that starts $i$ and $h$ the index within worker $g$'s start list (first $\sqrt{W}$ workers invoke $\frac{W-\sqrt{W}}{\sqrt{W}}$ others each). The \textit{startup} time indicates when the worker has started, and \textit{ready} time when the worker can start its data processing tasks (both since query epoch). This is different for some workers in the two-level invocation, as they must first invoke other workers---which is part of the billed time. 


\subsection{Base overhead}

Every worker performs a certain amount of base tasks that are independent of the workload, including initializing its variables, extracting input, allocating memory, and packaging its output. We model this (billed) overhead time as a single constant $T_{\vartext{base}}$.


\subsection{Process, compress, and network rates}

We model the process, compress, and network rate as three constants (dependent on the memory size chosen): $R_{\vartext{network}}$,  $R_{\vartext{compress}}$, and $R_{\vartext{process}}$. We consider network and compress rate as independent of the workload, whereas the process rate as dependent. These constants can be attained through measurement (as is done in \S\ref{sec:applied-model}), by applying some model (\eg using Amdahl's law) to account for the effect of more vCPU, or a combination of the two. Network rate is the rate at which both transfer of read and written data takes place.


\subsection{Input}
\label{sec:model-input}

The input is solely concerned with the reading and processing of the input:

$$T_{\vartext{input}} = max(\frac{D_{\vartext{input}}}{R_{\vartext{network}}}, \frac{D_{\vartext{input}}}{R_{\vartext{process}}})$$


\subsection{Exchange(s)}

The purpose of an exchange operator is to partition data such that each function receives all tuples with the same key. An exchange can be done in one or more levels, which for two or more levels means that first functions exchange in a group, and then having the groups exchange (and so forth depending on number of levels). An exchange operator is used to accomplish join and reduce operations, among others. Its operation consists of compressing and writing the data, and subsequently reading and processing the data destined to the function. The number of the group size in an exchange level is defined as $G_{\vartext{ex[j]}}$, a value we calculate based on the total number of workers and whether an exchange is one or two levels. We define an additional penalty factor for each worker participating in the exchange level, $T_{\vartext{overhead[j]}}$. For each exchange level $j$:

$$T_{\vartext{ex[j]}} = \frac{D_{\vartext{ex[j]}}}{R_{\vartext{compress}}} + \frac{D_{\vartext{ex[j]}}}{R_{\vartext{network}}} + max(\frac{D_{\vartext{ex[j]}}}{R_{\vartext{network}}}, \frac{D_{\vartext{ex[j]}}}{R_{\vartext{process}}}) + G_{\vartext{ex[j]}}\times T_{\vartext{overhead[j]}}$$


\subsection{Output}

The output is solely concerned with the amount of data to compress and subsequently written to storage:

$$T_{\vartext{output}} = \frac{D_{\vartext{output}}}{R_{\vartext{compress}}} + \frac{D_{\vartext{output}}}{R_{\vartext{network}}}$$


\subsection{Postprocessing}

The driver must receive the final results from the workers and process them to return a final result to the user. We model this as a constant: $T_{\vartext{postprocess}}$.


\subsection{Requests costs}

Unlike virtual machines, which almost exclusively communicate directly between each other, most serverless data processing systems communicate through storage services (\eg S3~\cite{starling, lambada}). In general, cloud providers charge for the access to the storage service. In exchanges between many participants, the number of requests can become a significant cost factor. When applying the advisory tool to a system, we assume we can give a reasonable estimate of the request cost $C_{\vartext{requests}}$.


\subsection{Overall Completion Time and Cost}

The completion time is defined as the last worker completing and its result having been collected by the driver:
$$T_{\vartext{completion}} = T_{\vartext{ready}}(W) + T_{\vartext{base}} + T_{\vartext{input}} \;\;+$$ 
$$(\sum_j T_{\vartext{ex[j]}}) + T_{\vartext{output}} + T_{\vartext{postprocess}}$$

The billable time of each worker does not directly equal the completion time. Instead, it equals the sum of time the worker was alive:
$$T_{\vartext{billable}} = \sum_i T_{\vartext{alive}}(i)$$

The alive (\ie billable) time of a worker is only independent from other workers if there no exchanges as else it has to pace with the slowest worker. We define the worker alive time of a simple scan (without exchanges) as:
$$T_{\vartext{alive}}(i) = T_{\vartext{ready}}(i) - T_{\vartext{startup}}(i) + T_{\vartext{base}} + T_{\vartext{input}} + T_{\vartext{output}}$$

Similarly, if there are exchanges, the workers have to wait until the last worker is ready. Hence, we define the worker alive time for a query with exchanges as:
$$T_{\vartext{alive}}(i) = T_{\vartext{ready}}(W) - T_{\vartext{startup}}(i) + T_{\vartext{base}} + T_{\vartext{input}} + \sum_j T_{\vartext{ex[j]}} + T_{\vartext{output}}$$

The final cost is the sum of the worker runtime cost and the sum of the requests cost:
$$C_{\vartext{total}} = T_{\vartext{billable}}\times \vartext{price/ms} + C_{\vartext{requests}}$$

\subsection{Model analysis}
\label{sec:general-model-analysis}

The general model has several parameters, each of which effects increase in completion time and/or financial cost, the latter either through increased billed serverless time or requests cost. The general tendencies are as follows. Firstly, increasing the number of workers ($W$) can have the following effects:

\begin{itemize}
    \item increased completion time as more workers must be invoked and exchange overhead is larger;
    \item decreased completion time as the amount of data each worker reads, writes and exchange reduces;
    \item increased cost through billed time as there is base overhead and invocation;
    \item and increased cost through request cost as there are more exchange members.
\end{itemize}

The above effects are under the assumption of invocation and exchange levels remaining the same. Increasing the worker memory size ($M$) can have the following effects:

\begin{itemize}
    \item decreased completion time as each worker can network send/receive, process and compress at a higher rate;
    \item increased cost through billed time as each time unit is more expensive.
\end{itemize}

As these examples show, several effects on completion time and financial cost compete against each other when increasing the number of workers ($W$) and memory size ($M$), increasing the intrinsic difficulty of estimating either. 


\section{Advisory tool and its evaluation methodology}
\label{sec:evaluation-methodology}

The advisory tool we propose uses the estimation model just presented. The tool aims to select a configuration balancing completion time and cost as there is no configuration (\ie number of workers $W$ and their memory size $M$) that minimizes both. In the next sections we describe how we define the best configuration (\S\ref{sec:pareto-knee}) and how we evaluate the advice (\S\ref{sec:advice-evaluation}).


\subsection{The Pareto knee}
\label{sec:pareto-knee}

\begin{figure}
 \begin{center}
  \includegraphics[width=0.5\textwidth]{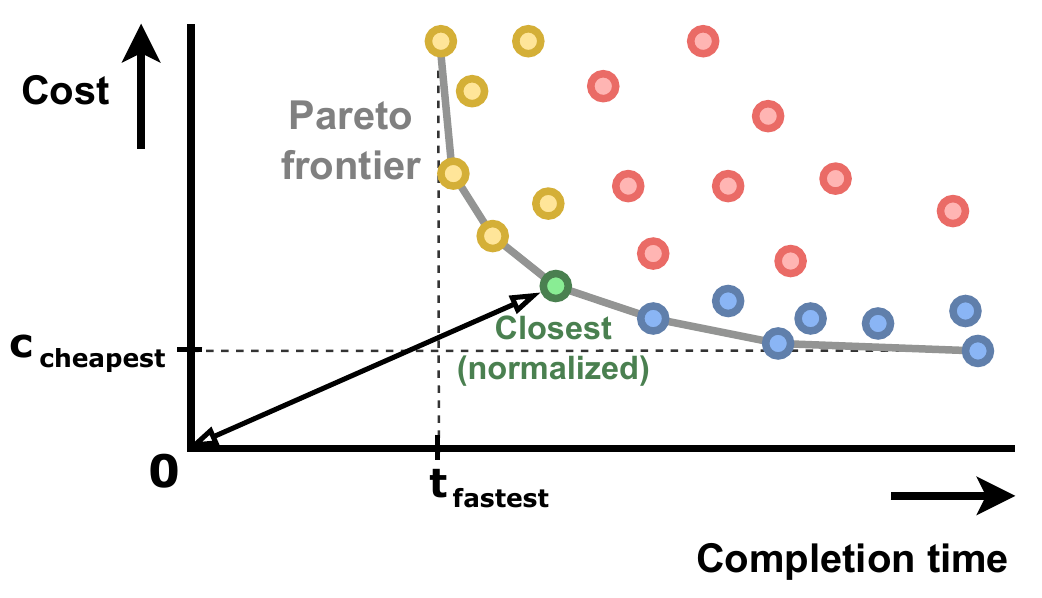}
  \caption{Pareto plot of completion time and cost. The circles are candidate configurations (the coloring corresponds to Fig.~\ref{fig:advice-outcomes}). The \textit{knee-opt} is the configuration closest (normalized) to the optimum of cheapest cost and fastest completion time.}
  \label{fig:pareto-knee}
  \vspace{-12pt}
 \end{center}
\end{figure}

The choice of configuration is a multi-objective optimization of completion time and cost, and as such can be visualized as a two-dimensional Pareto plot  (Fig.~\ref{fig:pareto-knee}). We consider as suitable configurations those closest to the "knee" of the Pareto frontier. We automate the picking of this "knee" using the following method.

Let $F$ be the set of candidate configurations. For each candidate configuration $i\in F$, we define its completion time as $t_i$ and its cost as $c_i$. We select as optimum the configuration with the shortest distance to the origin, with its completion time and cost normalized by the fastest and cheapest possible outcomes across all configurations:

$$t_{\vartext{fastest}} = min_{i\in F}\;t_{i}\;\;\;\;\;\;\;\;\;c_{\vartext{cheapest}} = min_{i\in F}\;c_{i}$$
$$d(i) = \sqrt{\alpha(\frac{t_i}{t_{\text{\textit{fastest}}}})^2 + \beta(\frac{c_i}{c_{\text{\textit{cheapest}}}})^2}$$
$$\text{\textit{knee-opt}} = arg\,min_{i\in F}\;d_{i}$$

The selection of this point is depicted visually in Fig.~\ref{fig:pareto-knee}. In our work, we set the weights $\alpha$ and $\beta$ to one, which means that we value completion time and cost equally. For example, a configuration with $2\times t_{\text{\textit{fastest}}}$ and $1\times c_{\text{\textit{cheapest}}}$ has the same score as a configuration with $1.6\times t_{\text{\textit{fastest}}}$ and $1.6\times c_{\text{\textit{cheapest}}}$ approximately.\footnote{As $\sqrt{\nicefrac{5}{2}}\approx 1.6$} A related work on multi-objective optimization for cloud data analytics proposed a similar method, but instead uses as metric the Euclidian distance directly to the idealized fastest and cheapest point (the so-called ``Utopia point'')~\cite{song2020boosting} (further discussed in \S\ref{sec:related-work}). We chose to use the normalized distance to the origin such that the resulting distance metric is relatively interpretable and comparable.


\subsection{Advisory tool and evaluation}
\label{sec:advisory-tool}
\label{sec:advice-evaluation}
The advisory tool uses the procedure above to select a configuration. As it does not know the actual completion times and costs, it instead uses the values supplied by the estimation model.

\begin{figure}
 \begin{center}
  \includegraphics[width=0.5\textwidth]{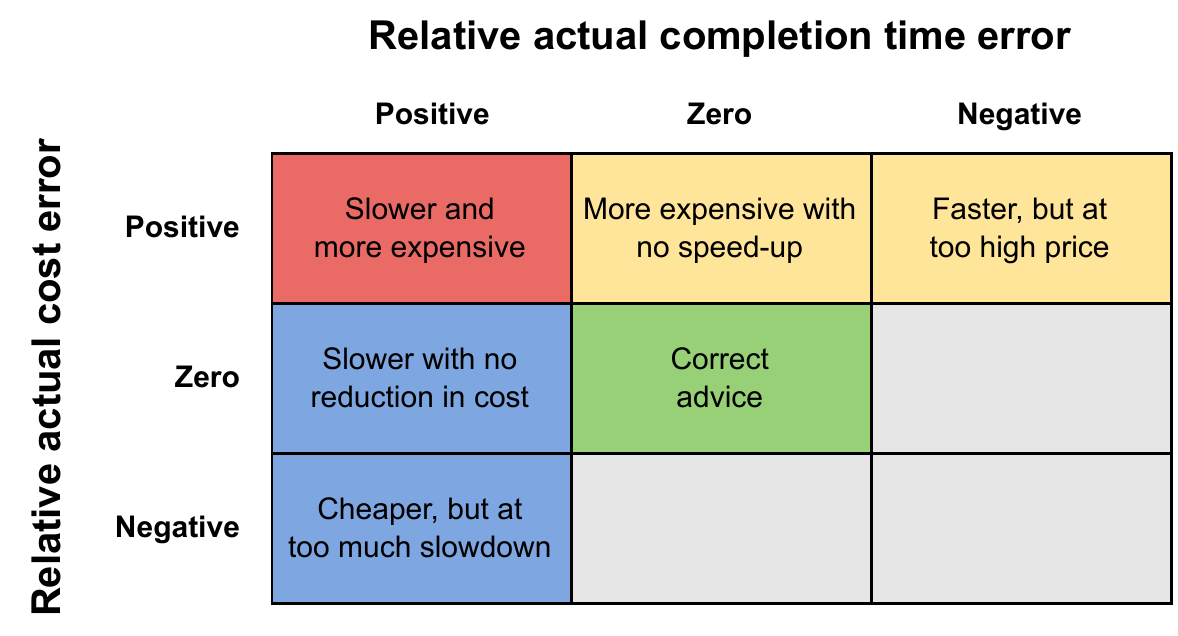}
  \caption{The advisory tool's possible outcomes. Because we compare actual outcomes, it is not possible to outperform the optimal configuration simultaneously in both dimensions as it is part of the Pareto frontier.}
  \label{fig:advice-outcomes}
  \vspace{-12pt}
 \end{center}
\end{figure}

In the later experiments, we run all configurations in the Cartesian product of two reasonable candidate lists of number of workers and memory size based on the partitioning and the query's operators' in-memory requirements, respectively. This provides us the data to determine the experimentally best choice of configuration. We use this as the baseline to compare the advice against. The primary comparison metric is the relative difference between their distance metric. Two additional metrics we use are the signed relative actual completion time and actual cost error of the advised configuration to the optimal configuration. If the advisor chooses the actual optimal configuration, both errors are zero. A negative relative error means that in that one dimension, the configuration selected was better. At most one of the two dimensions (time or cost) can have negative relative error as the optimal configuration is part of the Pareto frontier. The possible outcomes of an advice issued by the configuration advisor are shown in Fig.~\ref{fig:advice-outcomes}.

To illustrate, consider the following example. Suppose the best configuration has $t=1.1\times t_{\vartext{fastest}}$ and $c=1.3\times c_{\vartext{cheapest}}$, whereas the advised configuration outcome has $t=1.4\times t_{\vartext{fastest}}$ and $c=1.2\times c_{\vartext{cheapest}}$. The distances of the best and the advised configurations 1.70 and 1.84, respectively. The aforementioned metrics are in this case the following: the advised configuration is overall $\frac{1.84-1.70}{1.70}=8\%$ worse than the experimentally determined best choice and it runs $\frac{1.4-1.1}{1.1}\times100\% = 27\%$ slower at an $\frac{1.2-1.3}{1.3}\times100\% = -8\%$ cheaper cost.

\section{Applying the model: Lambada}
\label{sec:target-system}

To test the tool, we use it on Lambada~\cite{lambada}. We first describe Lambada (\S\ref{sec:target-system-description}), and then explain the experiments conducted (\S\ref{sec:target-system-measurement}). We apply the general model by filling in the Lambada/AWS-specific parameters (\S\ref{sec:applied-model}) and finally analyze the effect of the filled values in terms of overarching configuration preference (\S\ref{sec:applied-model-analysis}).


\subsection{Description}
\label{sec:target-system-description}

Lambada~\cite{lambada} takes as user input a computational DAG of operators, from which it devises a distributed query execution plan. The driver starts the selected amount and type of serverless workers (\ie AWS Lambda) and passes on the (distributed) plan to each worker. If many workers have to be started (if $W > 100$), the driver makes use of "two-level invocation": workers started first are also in charge of starting other workers. Once a worker is ready, it starts executing the plan: each worker reads its data partitions from cloud storage (\ie AWS S3). The partitioned tables are stored in Parquet format, to take advantage of compression and the ability to only download the columns relevant to the query. Each worker then goes through the same computations (albeit on different data), and amongst them perform exchanges when data should be shared (\eg join, aggregation). Exchanges take place through S3: each worker writes to a file the data to send to other workers. The exchanges can either be one-level ("all-to-all") or two-level (if number of workers $W > 32$). Two-level exchanges double the amount to read and write, while quadratically reducing the number of read and write requests (which is what the cloud provider charges for). Once done, a workers returns its result by putting them in a shared result queue which the driver monitors. The driver collects all results, performs some last operations, and returns the result to the user.


\subsection{Instrumentation}
\label{sec:target-system-measurement}

\noindent To capture runtime and cost information, we have augmented Lambada with the following instrumentation:

\begin{itemize}[leftmargin=10pt,itemsep=2pt,topsep=2pt]

    \item \textbf{Query completion time.} The completion time is measured by the driver from the moment it starts the workers until it has returned the final result to the user.
    
    \item \textbf{Query cost.} Each worker keeps track of the time from when it started until it returns the final result to the SQS queue, which constitutes the amount of billable time. Each worker counts the number of HEAD, GET, and PUT requests and logs them.

\end{itemize}

We use worker logs to keep track of request counts but also for post-mortem analysis: to know how long certain operations and functions within the workers take. This is vital to identify bottlenecks and the occurrence of stragglers. The logs are not used by the system itself, and as such the upload (by the workers) and download (by the driver) of worker logs are excluded from both the billable time of the workers and the final completion time.

For network stability reasons, we use an AWS instance as the driver rather than a machine in our local cluster. We chose a \texttt{m5a.2xlarge} instance as it is comparable to a high-end laptop with its 8~vCPU and 32~GiB of memory. We deploy our instance in the \texttt{eu-west-1} region, where we also store the S3 data used for the experiments and launch the workers.

In the running of (micro)benchmarks (\S\ref{sec:applied-model}, \S\ref{sec:into-practice}, \S\ref{sec:large-benchmark}), Lambada exhibited in a very small set of the runs internal errors that caused the run to fail (in a probabilistic fashion), among which: the lack of idempotency when handling SQS messages (\ie failing in the extremely rare case a message is received twice if a delete fails), SSL validation errors due to too many concurrent invocations (\ie the SSL request times out), and in some rarer cases there were certain (likely network-related) exceptions that were not caught. In the few cases in which any of these occurred, they were rerun.

There is another class of failure, which is to be expected: it is possible for Lambada to fail because a worker ran out-of-memory or worker start-up reaching a time-out limit of 20s. We consider a run for which this happens (one or more times among its repeats; generally it is for all instances) an infeasible configuration.


\subsection{Applied model}
\label{sec:applied-model}

In this section, we fill in the parameters of the general model~(\S\ref{sec:general-model}) except the data sizes, which we fill in on a per-query basis in the experiments (\S\ref{sec:into-practice}, \S\ref{sec:large-benchmark}). 

\parab{Invocation.} The rate and delay at which a system can invoke workers depends on the invocation implementation and the cloud. We use 128 concurrent threads for invocation on the driver, and 32 on a worker. We perform a microbenchmark in which we start 512 workers using only the driver or only the worker, under either COLD start\footnote{By re-deploying the function beforehand to ensure it is not cached.} and HOT start.\footnote{By performing another run with at least as many workers beforehand, which warms up the AWS lambda caches.} Tab.~\ref{tab:independent} shows the mean invocation rates and the mean 99th percentile invocation delay across three repetitions. The later experiments of this paper all are HOT started, in order to have the many runs finish in feasible time (as re-deployment to achieve coldness takes long and is rate-limited by AWS).

\begin{figure}
 \begin{center}
  \includegraphics[width=0.5\textwidth]{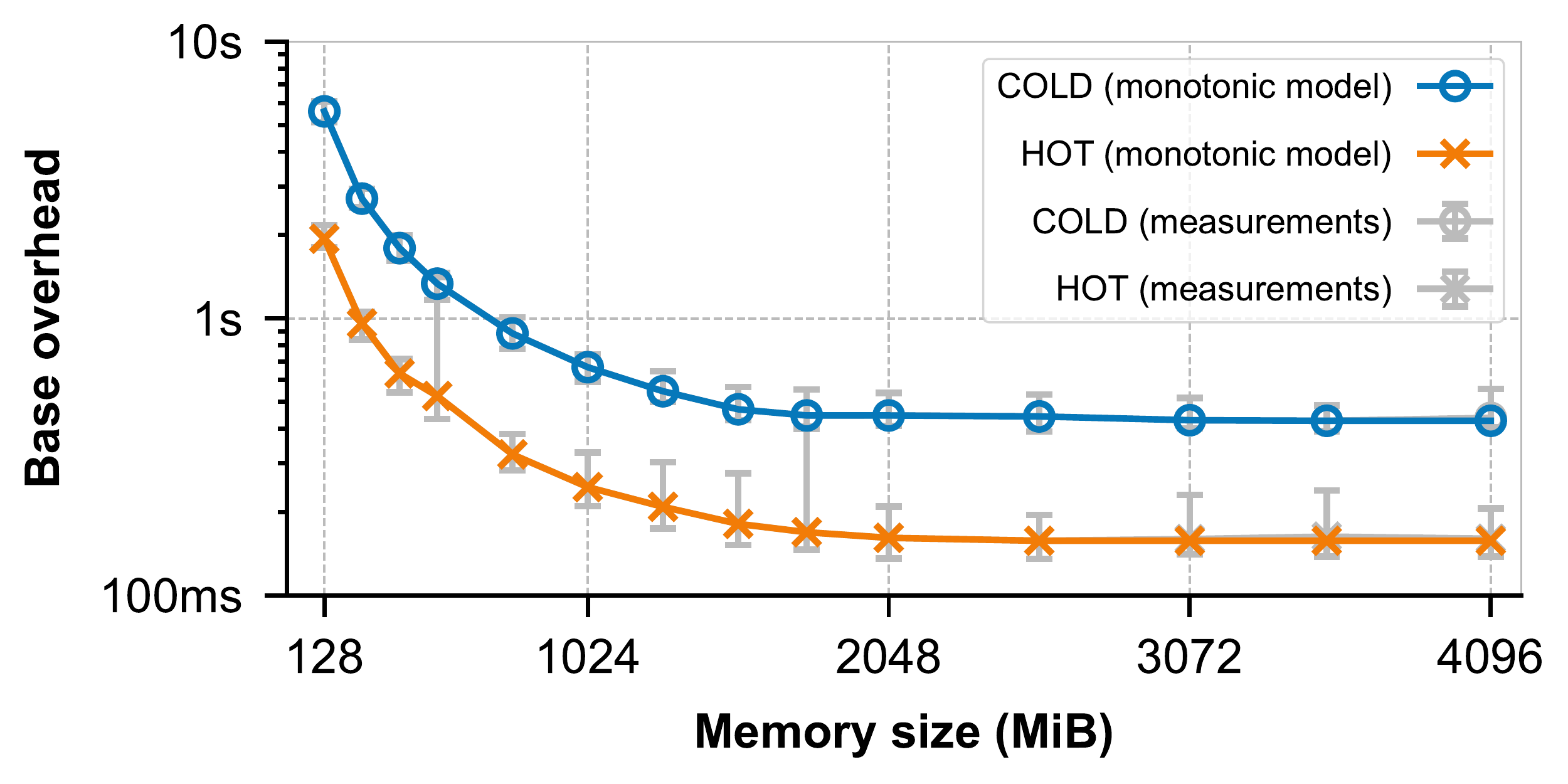}
  \caption{Base overhead. Error bars are min-max.}
  \label{fig:base-overhead}
  \vspace{-6pt}
 \end{center}
\end{figure}

\parab{Base overhead.} We perform a simple benchmark in which each worker retrieves a single small file. We subtract the scan duration from the total time the worker is deducted from its alive time. The result, its mean across three repetitions, is made monotonic, and is shown together with the measurements in Fig.~\ref{fig:base-overhead}.

\begin{figure}
 \begin{center}
  \includegraphics[width=0.5\textwidth]{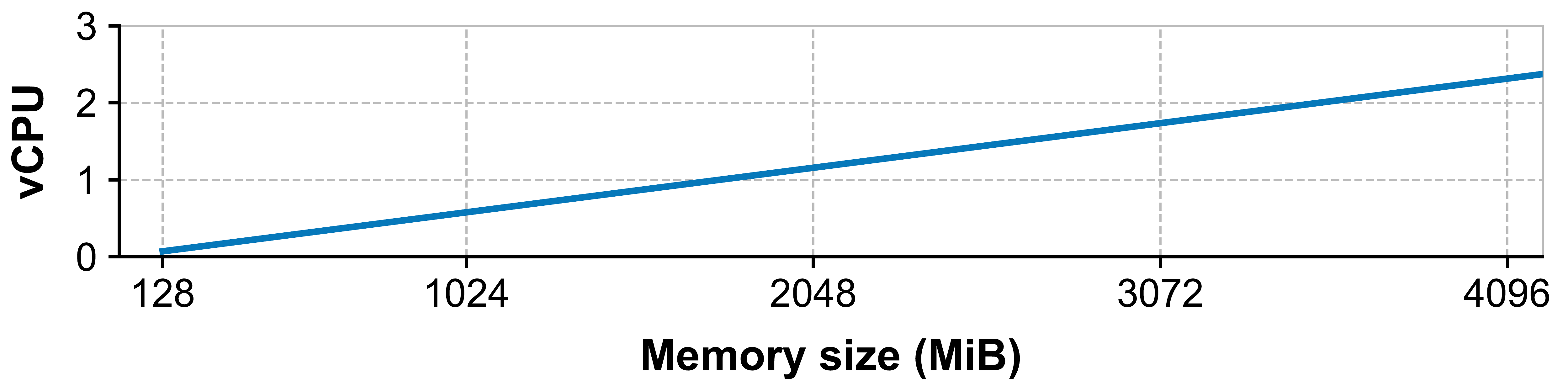}
  \caption{vCPU allocation on AWS (linear scaling with 1769~MiB/vCPU up to 10240 MiB).}
  \label{fig:vcpu}
  \vspace{-6pt}
 \end{center}
\end{figure}

\begin{figure}
 \begin{center}
  \includegraphics[width=0.5\textwidth]{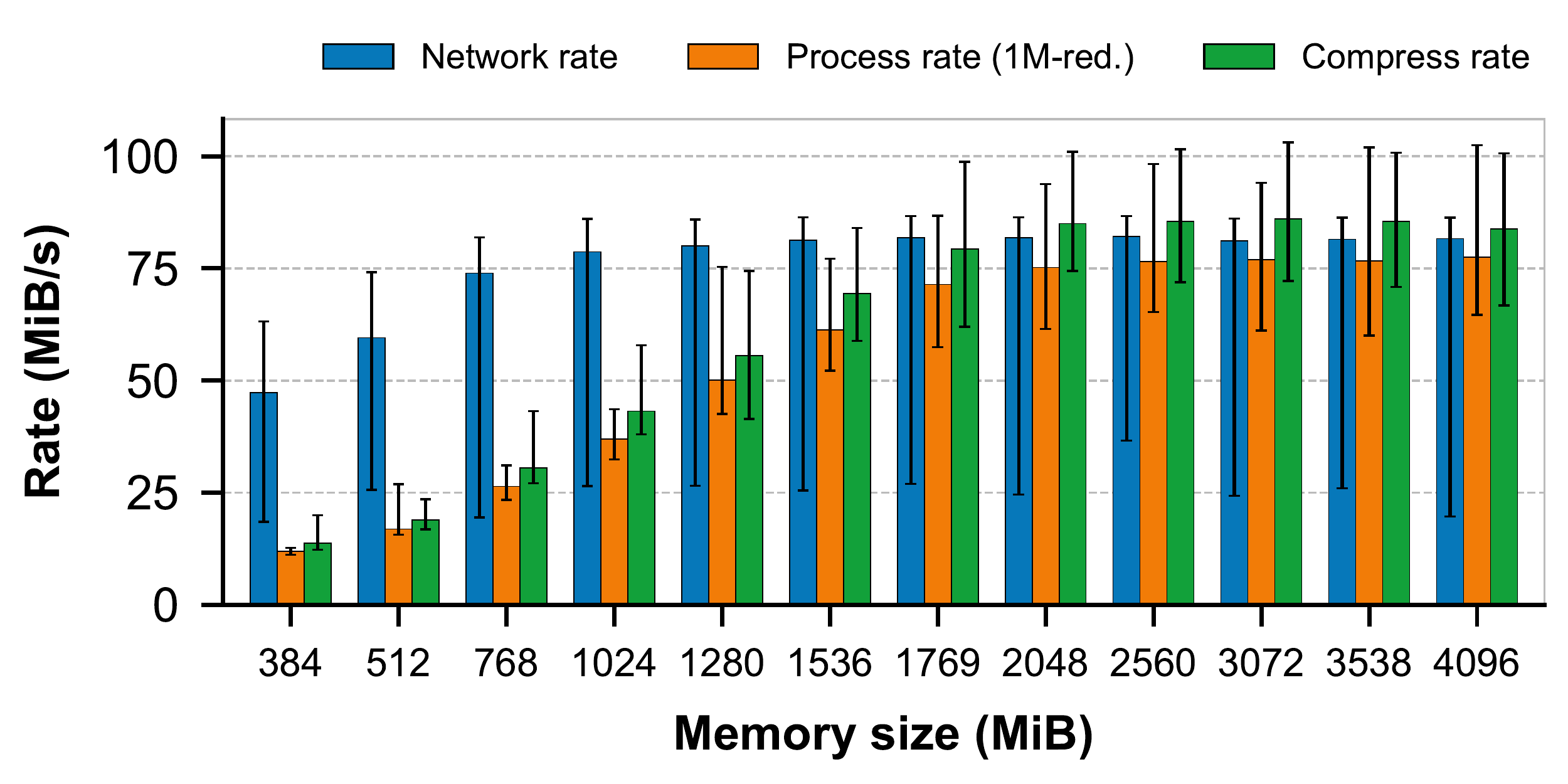}
  \includegraphics[width=0.5\textwidth]{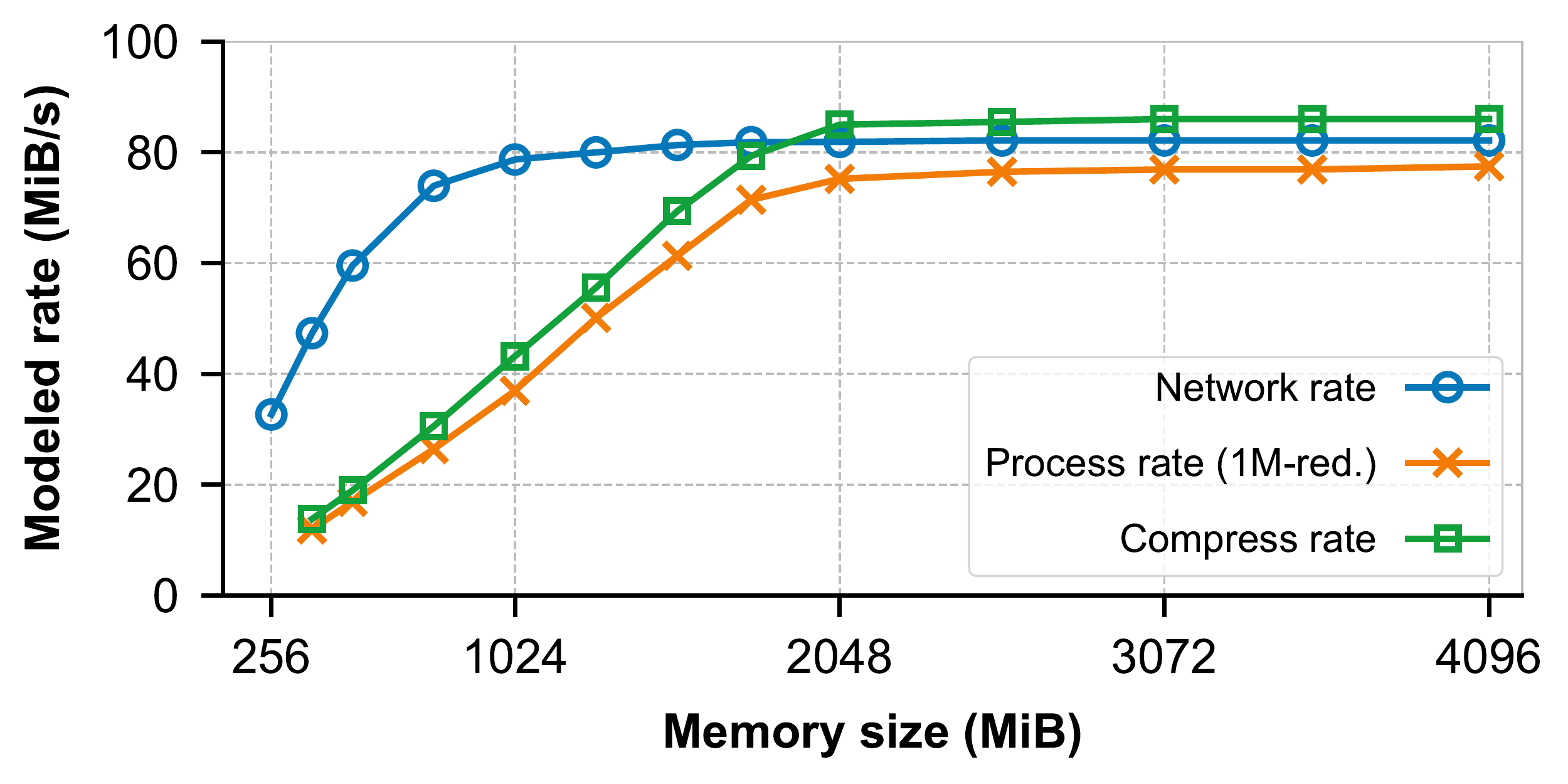}
  \caption{Network, process and compress rate (top: measurements, bottom: monotonic model). Error bars in top bar plot are min-max.}
  \label{fig:rates}
  \vspace{-6pt}
 \end{center}
\end{figure}

\parab{Compress rate.} Before writing data to storage, it must first be compressed into Parquet format. We assume the compression rate is independent of the type of underlying data. We perform an exchange microbenchmark with each of 32 workers reading and compressing \SI{32}{\mebi\byte} with two columns. The experiment outcome is shown in Fig.~\ref{fig:rates}(top). The mean compression rate (for all 96 samples across three repetitions) made monotonic is used in the model, depicted in Fig.~\ref{fig:rates}(bot). It scales proportionally with vCPU allocation (Fig.~\ref{fig:vcpu}) (single-core bound).

\parab{Network rate.} Unlike VMs~\cite{leis2021towardscostoptimal}, there is no network bandwidth info available for serverless functions (see Tab.~\ref{tab:cloud-offering}). Although it is observed that there is an initial higher network burst rate~\cite{lambada,boxer}, we model network rate at its sustained state. In the microbenchmark, each worker reads in a \SI{128}{\mebi\byte} file with eight columns. The mean network (for all 96 samples across three repetitions) made monotonic is used in the model, depicted in Fig.~\ref{fig:rates}(bot). Note that especially network rates can have outliers, we have observed very low rates every now and then (see minima in Fig.~\ref{fig:rates}(top)). Lambada has a small multi-threaded behavior in which, if it has two partitions to read in, it does so simultaneously. If it has sufficient compute available to it (at least more than \SI{768}{\mebi\byte}), it can achieve (temporarily) a network rate higher than the model predicts. We consider these conditions too system-specific and as such are not captured in the general model (in this microbenchmark, there is only a single input partition per worker).

\begin{table}[t]
    \centering
    \renewcommand{\arraystretch}{1.3}
    \begin{tabular}{| p{4.1cm} | C{3.6cm} | } \hline
        \textbf{Parameter} & \textbf{Value} \\ \hline \hline
        Driver invocation rate $R_{\vartext{1-inv}}$ & 142.3~inv/s \\ \hline
        Worker invocation rate $R_{\vartext{2-inv}}$ & 93.6~inv/s \\ \hline
        Invocation delay $T_{\vartext{inv. delay}}$ & COLD:~1263.4~ms, HOT:~677.7~ms \\ \hline
        Postprocess duration $T_{\vartext{postprocess}}$ & 188.5~ms \\ \hline
        Overhead per exchange member $T_{\vartext{overhead[j]}}$ & Scales linearly with the number of GET and HEAD requests\\ \hline
    \end{tabular}
    \caption{\em Memory-independent cloud-/system-parameters.}
    \label{tab:independent}
    \renewcommand{\arraystretch}{1}
    \vspace{-15pt}
\end{table}

\parab{Requests and exchange overhead.} Lambada fetches the exact columns it needs from each Parquet file. With the knowledge of tables/columns involved for each input scan and exchange, we can calculate the exact number of GET requests. Due to availability polling during exchanges, the number of HEAD requests can vary between runs. An exchange microbenchmark yielded approximately 3 HEAD requests for each exchange file. 
We use the number of requests and their duration to determine the exchange overhead. Microbenchmark measurements indicate a HEAD and a GET request take approximately \SI{13.1}{\milli\second} and \SI{18.4}{\milli\second} on average, though there were values of up to \SI{215}{\milli\second} (HEAD) and \SI{530}{\milli\second} (GET). The exchange overhead per group member $T_{\vartext{overhead[j]}}$ is set to the multiplication of the requests to read its exchange file and the request type duration. From the AWS documentation, we directly retrieve pricing for serverless function time (0.0000166667~\$ per GiB-second)~\cite{aws-lambda-pricing}, and read and write requests (respectively 0.0004~\$ and 0.005~\$ per 1000 requests)~\cite{s3-pricing}.

\parab{Post-processing.} We have 128 workers each fetch a file of 32~MiB, and measure the duration between the last worker having returned its result and the driver finishing. Tab.~\ref{tab:independent} shows the mean duration across 10 repetitions.

\parab{Processing rate.}  The process rate of a query is the most difficult parameter to estimate, as it is query dependent. As an admittedly imperfect solution, we use the reduction to 1 million unique values as the process operation; its internal data structure, the hash table, is used across many common operations such as joins, reductions, and group-by's. The results are shown in Fig.~\ref{fig:rates}(top). Note that Lambada's operator execution model is single-threaded and only marginally benefits from multiple vCPUs: it roughly scales proportionally with vCPU allocation (Fig.~\ref{fig:vcpu}) up until single core.
 

\subsection{Applied model analysis}
\label{sec:applied-model-analysis}

\noindent The general model applied to Lambada (\S\ref{sec:applied-model}) preserves the general model influences (\S\ref{sec:general-model-analysis}) by monotonically modeling the rates and overhead. Its modeled sustained network rate is capped out at \SI{82.1}{\mebi\byte\per\second}, with \SI{78.7}{\mebi\byte\per\second} being already reached at $M=1024$ (0.58~vCPU). In addition, due to its mostly single-threaded nature, its process rate and compress similarly are capped at respectively \SI{77.5}{\mebi\byte\per\second} and \SI{86.0}{\mebi\byte\per\second}, with \SI{75.2}{\mebi\byte\per\second} and \SI{85.0}{\mebi\byte\per\second} being already reached at $M=2048$ (1.16~vCPU). This means that:

\begin{itemize}
    \item if network is the bottleneck, increasing the memory size beyond $M=1024$ has little effect on completion time;
    \item if compute is the bottleneck, increasing the memory size beyond $M=2048$ has little effect on completion time.
\end{itemize}


\section{Into practice}
\label{sec:into-practice}

In this section, we describe the manner how to apply the model and the practical guidelines in terms of choosing the correct configuration that it implies. We first describe in \S\ref{sec:evaluation-practice} how we apply the evaluation methodology of the advisor (\S\ref{sec:evaluation-methodology}) in our benchmarks. In the context of serverless data processing tasks, we characterize workloads into two categories: (1) workloads that solely scan data (\S\ref{sec:scan}), and (2) workloads that involve one or more exchanges (\S\ref{sec:exchange}). Either category has its own trend in the sweet spot in its choice of number of workers and memory size. This section lays the groundwork for the large scale benchmark in \S\ref{sec:large-benchmark}.

\begin{figure*}[t]
	\centering
 	\hfill
    \begin{subfigure}[b]{0.4\textwidth}
        \centering
		\includegraphics[width=\textwidth]{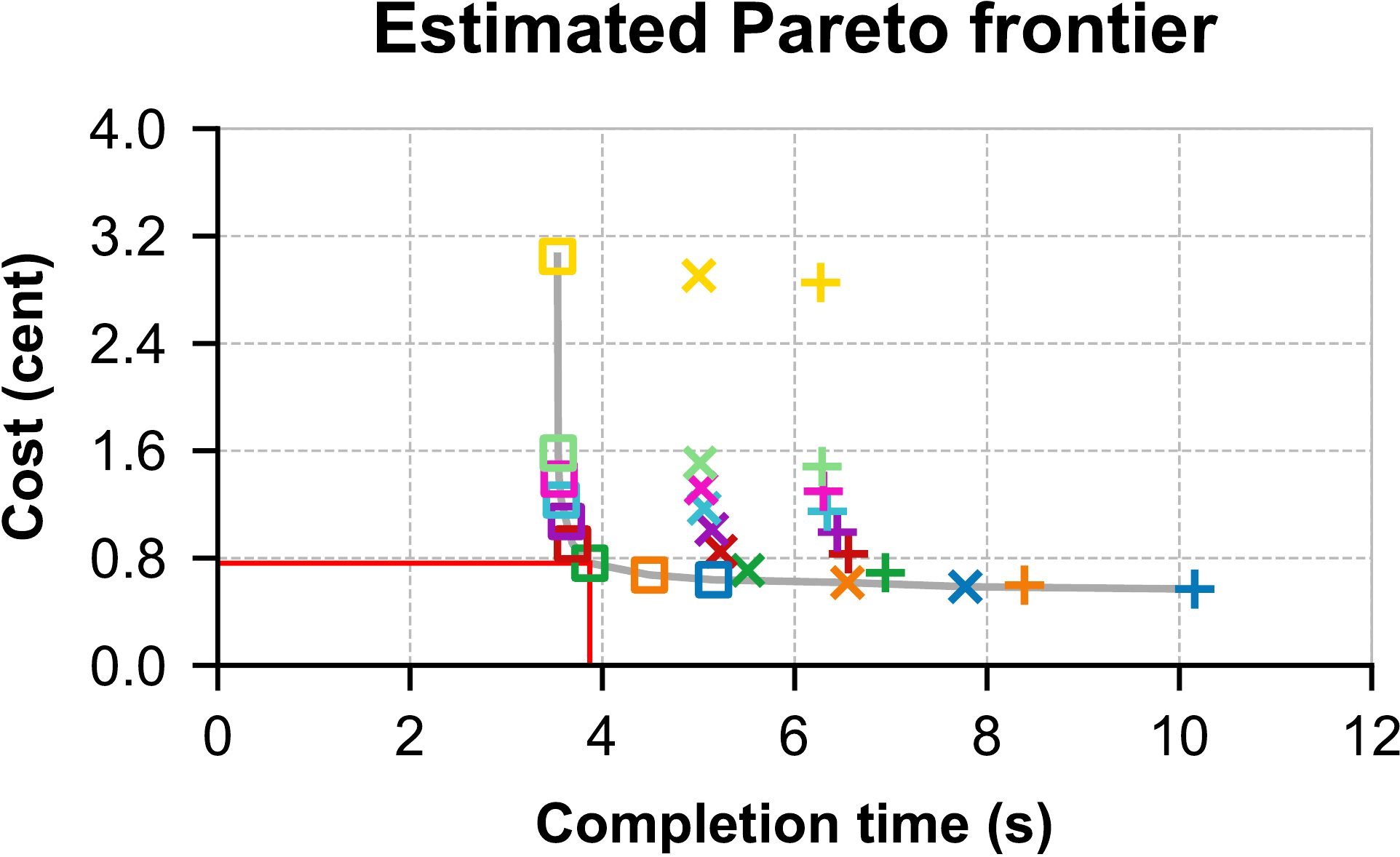}
		\caption{}
		\label{fig:scan-estimated}
    \end{subfigure}
 	\hfill
    \begin{subfigure}[b]{0.4\textwidth}
        \centering
		\includegraphics[width=\textwidth]{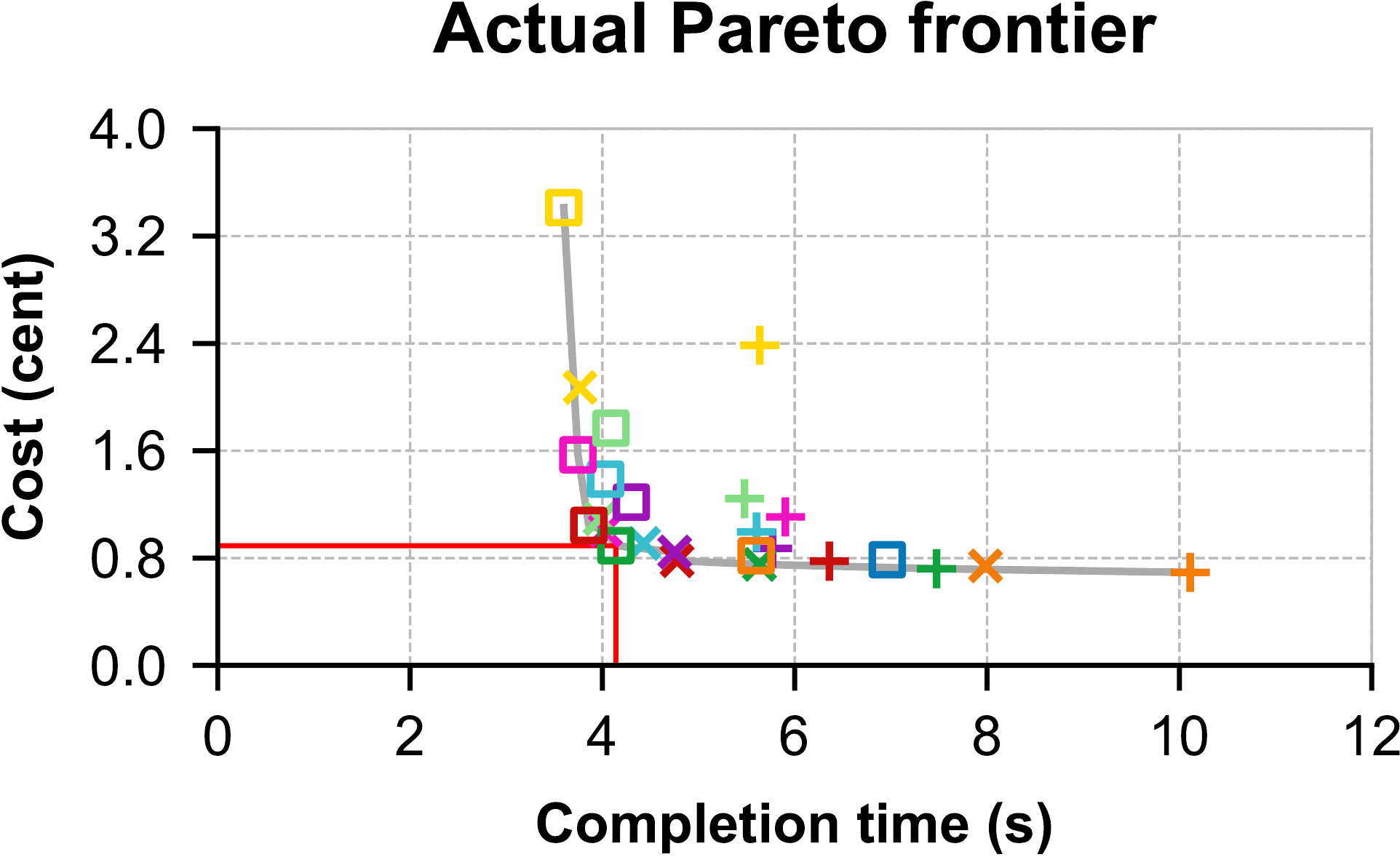}
		\caption{}
		\label{fig:scan-actual}
    \end{subfigure}
 	\hfill
    \begin{subfigure}[t]{0.19\textwidth}
        \centering
        \raisebox{1.0cm}{
        \includegraphics[width=\textwidth]{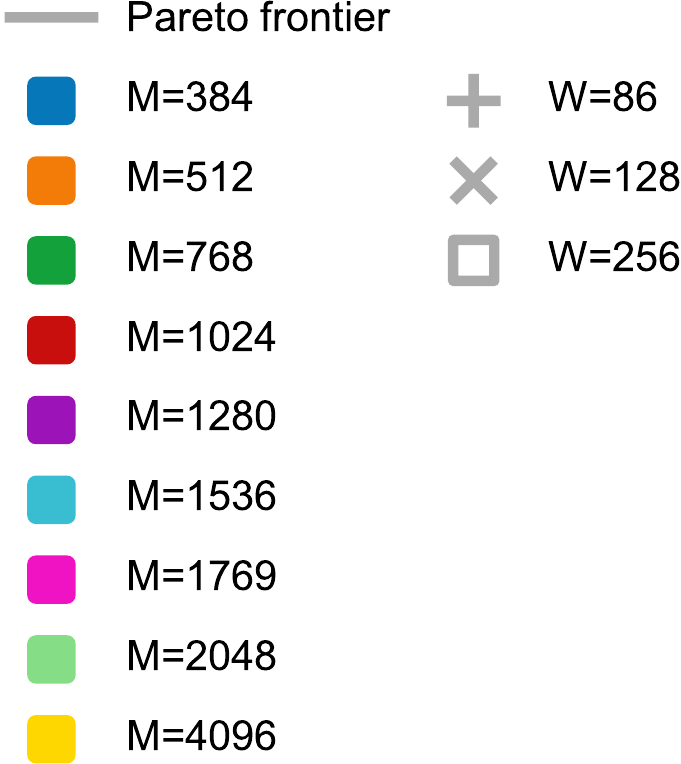}
        }
    \end{subfigure}
	\vspace{-18pt}
	\caption{Scan benchmark: Pareto frontiers with the red lines pinpointing their kneepoint (estimated = actual: W=256, M=768).}
	\label{fig:scan}
	\vspace{-6pt}
\end{figure*}


\subsection{Methodology}
\label{sec:evaluation-practice}

The goal of our work is to characterize serverless data processing performance in general, which implies some version of ideal circumstances. As such, we want to evaluate the model's rough estimation and advice using an ideal system. In particular, this means we do not wish to compare to runs that are unfortunate enough to encounter stragglers (\S\ref{sec:prediction-challenges}), which are in particular present for network rates and requests (\S\ref{sec:applied-model}, \S\ref{sec:applied-model-analysis}). We similarly can only perform a limited number of repeated runs with a Cartesian product of memory sizes and number of workers. In order to eliminate outliers, we opt to choose the run best suited to be a kneepoint. We choose this by first performing the distance metric using all data points, and for each configuration choose the run which had the smallest distance metric. The chosen configuration is then the experimental best and used as a reference for error calculation. As such, each point in any plotted actual Pareto frontier (\eg Fig.~\ref{fig:scan-actual}) is not the mean across all repetitions, but the best run among them. Note that although we do choose the representative data point for each configuration to be the best, the normalization is done with the cheapest and fastest outcome across all repetitions. This is done because the representative data point selection is to account for outliers, whereas normalization should be done with the true lowest values. The theory as presented in \S\ref{sec:advice-evaluation} is upheld, as the fastest and cheapest value can only be lower by considering all repetitions.

It is possible that the advisor recommends a configuration which does not have an actual experimental outcome (\ie due to running out of memory or hitting the timeout limit), as it is not always aware of this. In this case, which is exceedingly unlikely occur, the advisor error would be set to infinite -- though, in none of our experiments this was the case as these are generally the extremes.


\subsection{Scan workloads}
\label{sec:scan}

In a \textbf{scan workload}, there is no communication between workers. Workers read in the input partitions assigned to them, scan through their assigned data, and return their result directly to storage or the driver in case of simple aggregation. The typical use case for this workload type is to select tuples that fit certain criteria, or very simple aggregations whose return value is a few tuples (\eg counting, summing, reduction to very few tuples). The associated computational requirement is as such not much. The primary bottleneck is their reading of input over the network.

Based on the applied model analysis (\S\ref{sec:applied-model-analysis}), around a memory size of 768-1024~MiB there starts to be diminishing returns on network rate when increasing further. Because the workers do not exchange data in a scan, there is no strong penalty on the increase of workers to large numbers. Only the relatively minor base overhead and invocation increase the billed runtime, and only the latter to a small extent the completion time (as more workers must be started).

To study the scan workload, we perform the following experiment. We make use of a 32~GiB table of eight 64-bit columns filled with random values. It is partitioned into 256 Parquet files of 128~MiB each. The query counts the number of rows (which is $2^{29}$). As this operation is computationally simple, we set the model $R_{\vartext{process}}$ to a value larger than $R_{\vartext{network}}$, which means it is of no consequence. As only the large number of workers are of interest, we vary the number of workers $W\in\{86, 128, 256\}$ (corresponding to 3, 2, and 1 partitions/worker respectively). We vary the memory size at steps between 384~MiB and 4096~MiB. The configurations (W=86/128, M=384) did not make the 20 second start-up time limit and as such did not finish. Decreasing the number of workers further results in too long completion time, and decreasing the memory size further leads to out-of-memory (\ie infeasible) and too slow completion. Both of which are not part of the regime of interest, which is the knee of the curve: achieving decent performance at a cost-efficient price point. Each configuration run is repeated five times.

\begin{figure*}[t]
	\centering
 	\hfill
    \begin{subfigure}[b]{0.4\textwidth}
        \centering
		\includegraphics[width=\textwidth]{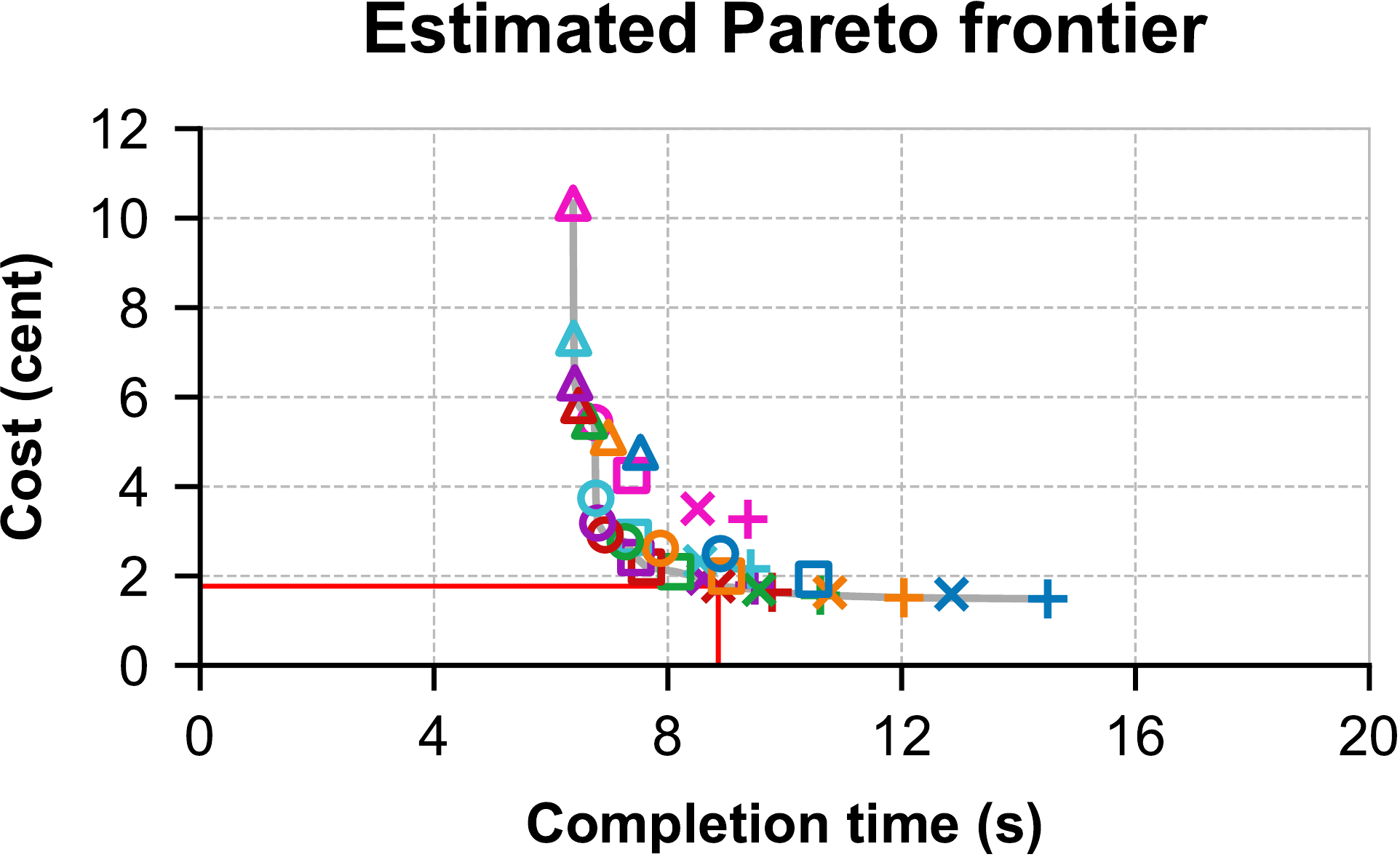}
		\caption{}
		\label{fig:exchange-estimated}
    \end{subfigure}
 	\hfill
    \begin{subfigure}[b]{0.4\textwidth}
        \centering
		\includegraphics[width=\textwidth]{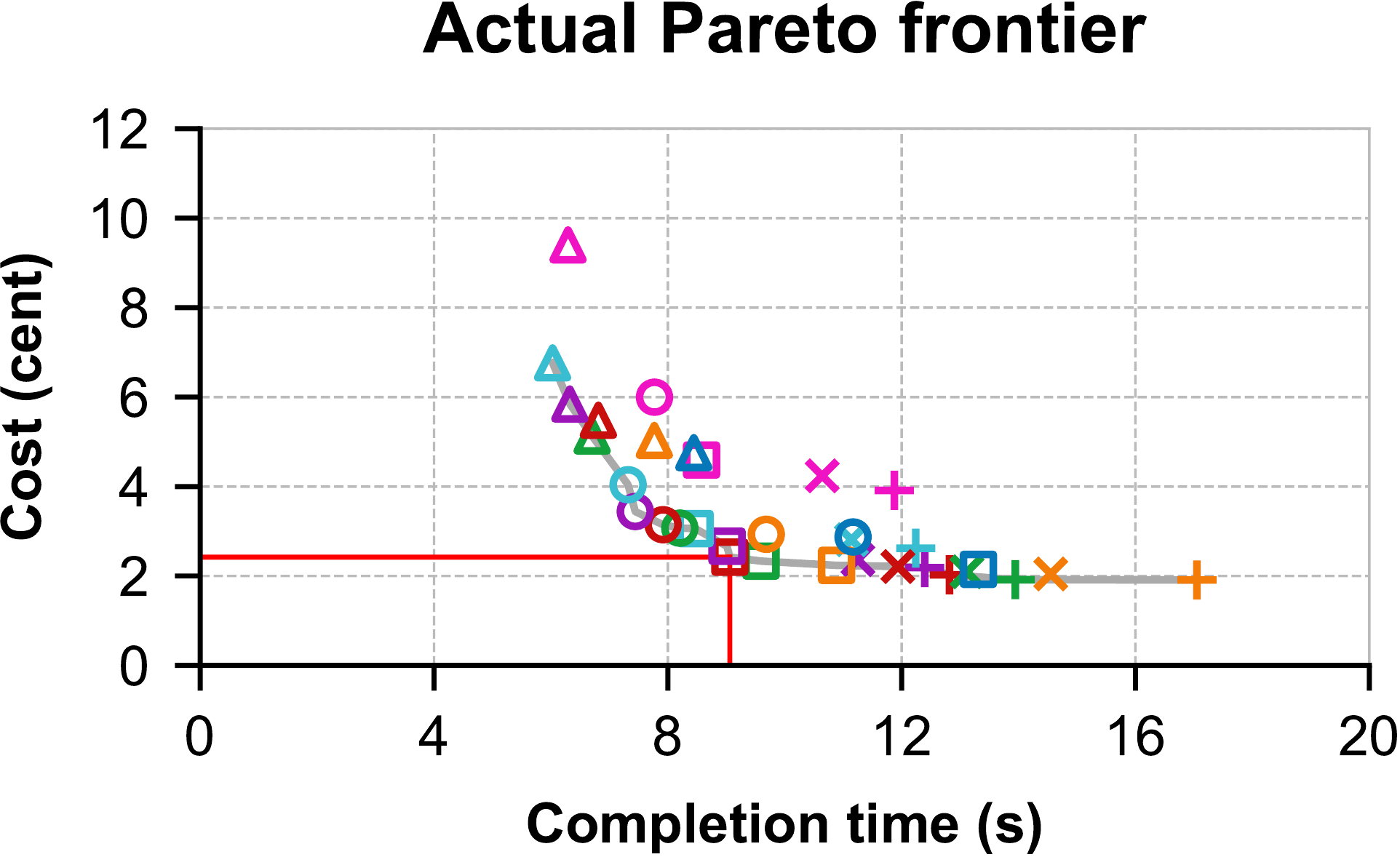}
		\caption{}
		\label{fig:exchange-actual}
    \end{subfigure}
 	\hfill
    \begin{subfigure}[t]{0.19\textwidth}
        \centering
        \raisebox{1.52cm}{
        \includegraphics[width=\textwidth]{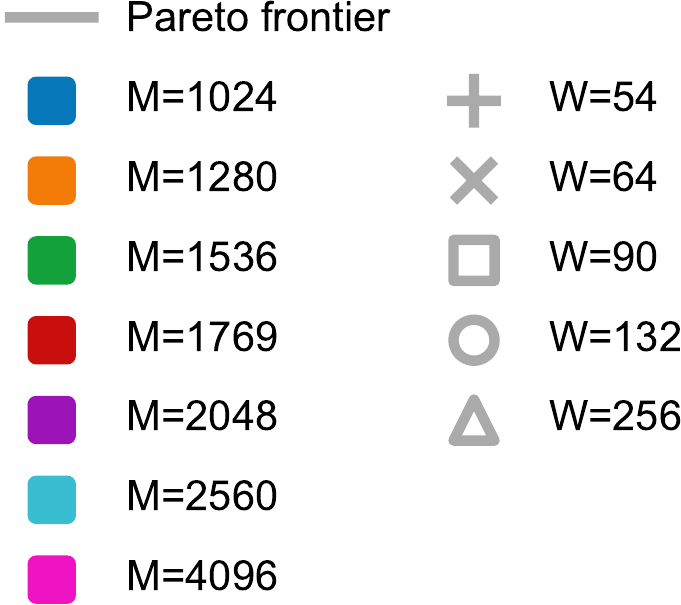}
        }
    \end{subfigure}
	\vspace{-18pt}
	\caption{Exchange benchmark: Pareto frontiers with the red lines pinpointing their kneepoint (estimated: W=64, M=1769, actual: W=90, M=1769).}
	\label{fig:exchange}
	\vspace{-6pt}
\end{figure*}

The estimated Pareto frontier is depicted in Fig.~\ref{fig:scan-estimated}: the maximum number of workers $W=256$ with a modest memory size of $M=768$ is advised to achieve both good completion time and cost. This advice matches the experimentally determined best choice, as is seen in the actual Pareto frontier in Fig.~\ref{fig:scan-actual}

Note that the estimation does not always match the actual outcome. For $W=256$ the completion time and cost estimate is slightly lower: this is due to regular variance of network rate. For $W=128$ and $W=86$ the estimates are consistently too high in completion time and cost for memory sizes over 768~MiB: this is because $W=128$ and $W=86$ have more than one input partition, such that Lambada can download two input partitions simultaneously. With its dual download and sufficient compute to it, it can make use of the temporary burst rate serverless functions on AWS have available to it~\cite{lambada}. Our model asserts only a sustained rate, which purposefully does not capture such system-specific intricacies.

These results show that scan workloads are generally network-intensive, not compute-intensive (unless the scan involves a computationally heavy operation \eg a reduction-by-key), and do not include exchanges. Their optimal configuration is to make use of many workers with a memory size that provides high network bandwidth.


\subsection{Exchange workloads}
\label{sec:exchange}

In an \textbf{exchange workload}, workers read in their respective input data, and then proceed to perform one or more exchanges to facilitate certain global operations such as joining, grouping, or sorting. The associated computation can be considerably intensive, and its complexity can depend on the data~(\S\ref{sec:data-processing-limits}). An exchange workload has to not only read input data and write output data, but also read and write for each exchange. As such, it is network-intensive from a rate perspective. Besides rate limits, there is also a cost to an increase in exchange members: there is a limit to the amount of outstanding requests and ongoing connections. This is particularly important for Lambada, which communicates through the writing and reading of files to storage (a practice not unique in serverless data processing systems~\cite{starling}) as inter-lambda communication is not (easily~\cite{boxer}) possible. For every exchange member, a worker has to read in all columns individually.

Following the applied model (\S\ref{sec:applied-model}), we expect the memory size with near-highest network rate, process rate, and compress rate to be the best pick: around $M$=1769-2048 appears to be the best. We expect a number of workers that balances for completion time the decreasing influence of more workers processing the amount of data (\ie rate) and the increasing influence of the extra overhead caused by more exchange members. Another factor we must consider is out-of-memory: hash tables can grow to large sizes with many unique keys for instance.

We examine the exchange workload through a benchmark. We use a 4~GiB table of one column filled with 64-bit random values, partitioned into 256 Parquet files of 16~MiB each. The query is to count the number of unique rows (which is in expectation $2^{29}$) -- which requires a reduction operation. This is a rather compute-intensive operation: every tuple processed will be inserted into the hash table that facilitates the reduction. We take the 1M-reduction ($\approx2^{20}$) processing rate from Fig.~\ref{fig:rates}, although the operation will likely be slower. This is one of the fundamental limitations of estimating arbitrary compute: it is as arbitrarily difficult. We vary the memory size between 1024 and 4096~MiB in several steps. We choose a number of workers in the mid-range, with 1 to 5 partitions/worker: $W\in\{54, 64, 90, 132, 256\}$. Reducing memory size or number of workers further results in out-of-memory cases in the Cartesian product. The configurations (W=54/64, M=1024) already ran out-of-memory and as such did not finish. Some of the number of workers (54, 90, 132) are slightly higher than the partitions/worker requires: this is to have a performant two-level exchange as we must divide the workers in two groups $X, Y$ such that $X\times Y=W$ exactly while minimizing $X+Y$ to get good performance.

The estimated Pareto frontier (Fig.~\ref{fig:exchange-estimated}) generally underestimates the completion time and cost when compared to the actual Pareto frontier (Fig.~\ref{fig:exchange-actual}). This is caused by the unexpectedly slower processing experienced in practice. Nevertheless, what is important for the advisor is that the shape, the relative positioning of configurations to each other in completion time and cost, is preserved. The advisor recommends the ($W$ = 64, $M$ = 1769) configuration as best, whereas the experimentally determined best choice actually was the ($W$ = 90, $M$ = 1769) configuration. If the user would have followed the advice (as in practice, the user would not explore the space), there would have been a 32\% increase in completion time with a -9\% cost reduction. With additional prior information (\eg if the query is run repeatedly), this discrepancy could be adjusted and the estimation (and as such, the advice) improved. For example, by simply reducing the model process rate by one-third for all memory sizes in Fig.~\ref{fig:rates}, the advisor would elongate along the x-axis (completion time) and correctly pick the best configuration.

Unlike for scanning, an increase in workers leads also to an increase in the number of requests ($\BigO(W\sqrt{W})$~\cite{lambada}) and exchange overhead. These two factors cause the curve to go up with the number of workers: for instance, at ($W$ = 64, $M$ = 1769) a mere 15\% of cost is requests, whereas at ($W$ = 256, $M$ = 1769) it is 38\%. Note that the benchmark exchange has a large amount of exchanged data (its entire input). In other exchanges which exchange little data, throughput is likely to be less of a bottleneck, whereas the exchange overhead can dominate the completion time.

In exchange workloads, the optimal configuration makes use of a number of workers that balances total rate and the overhead/cost of the enlarged exchange. Exchange workloads are generally both network- and compute-intensive, and as such favors a memory size which offers both at reasonable cost.


\section{Large benchmark}
\label{sec:large-benchmark}

A version of the TPC-H benchmark~\cite{tpc-h} that runs on Lambada is used to explore how well the estimation performs for representative workloads~\cite{lambada}. Here we expect to see larger errors in estimating running time and cost but what matters is how close the suggested configuration is to the optimal. 


\subsection{Dataset generation}

\begin{table*}[t]
    \centering
    \begin{tabular}{| p{0.5cm} | p{1.35cm} | p{1.2cm} | p{3.2cm} || C{1.3cm} | C{1.3cm} | C{1.3cm} | C{1.3cm} | C{1.3cm} | C{1.3cm} | } \hline
        \textbf{SF} & \textbf{Partitions} & \textbf{Total size} & \textbf{Number of workers range} & \textbf{LINE- ITEM} & \textbf{ORDERS} & \textbf{PART- SUPP} & \textbf{CUS- TOMER} & \textbf{PART} & \textbf{SUP- PLIER} \\ \hline
        20 & 39 & 6.5~GiB & 2, 3, 5, 10, 13, 20, 39 & 109.6~MiB & 29.2~MiB & 21.7~MiB & 6.3~MiB & 3.4~MiB & 0.4~MiB \\ \hline
        100 & 192 & 33.0~GiB & 6, 12, 24, 48, 64, 96, 192 & 113.4~MiB & 30.4~MiB & 22.1~MiB & 6.4~MiB & 3.4~MiB & 0.4~MiB \\ \hline
        200 & 384 & 67.0~GiB & 12, 24, 48, 96, 128, 192, 384 & 114.2~MiB & 32.1~MiB & 22.1~MiB & 6.4~MiB & 3.4~MiB & 0.4~MiB \\ \hline
        500 & 960 & 168.7~GiB & 30, 60, 120, 240, 320, 480 & 115.0~MiB & 32.6~MiB & 22.1~MiB & 6.4~MiB & 3.4~MiB & 0.4~MiB \\ \hline
    \end{tabular}
    \caption{\em Large benchmark datasets of different scale factor (SF) used in the experiments: for each table, the measured mean Parquet partition file size is stated. REGION and NATION are omitted as their (small constant) size does not increase with SF.}
    \label{tab:tpc-h-size}
    \vspace{-16pt}
\end{table*}

For each scale factor (SF), we performed the following steps to prepare the dataset for consumption by lambda workers. (a) We generated the raw text CSV files using the Lambada~\cite{lambada} version of dbgen~\cite{tpc-h-dbgen}. (b) We use LINEITEM as the base to decide how to split all tables as it is the largest table. At SF=1, the size of LINEITEM is approximately 586.2~MiB uncompressed. After compression into a Parquet file, it is 193.3~MiB. We set the divisor to achieve a size of $\pm~100$~MiB per LINEITEM partition. We split each single large CSV file (of the eight tables) into the smaller files. (c) We convert each raw text CSV to Parquet format using Apache Arrow~\cite{apache-arrow}. The final result for a subset of the scale factors is depicted in Tab.~\ref{tab:tpc-h-size}. (d) The files are stored on AWS S3. Note that in query execution, only the relevant columns are read: a worker retrieving a partition thus only retrieves a fraction of the total size of each file. We did not sort the CSV files before splitting, so the min/max column metadata in the Parquet files does not allow reads to be skipped.


\subsection{Query modelling}

For each query, the model workload parameters must be determined to facilitate the rough estimation. 
The \textbf{selectivity} in the data flow we manually determine by loading the benchmark data (at SF=1) into a relational database, and manually executing selection queries corresponding to each phase (input, exchange(s), output). For example for Q4, approximately 63.2\% of the LINEITEM tuples and 3.8\% of ORDERS tuples pass the input selection criteria. Q19 of TPC-H was altered using the correct REG AIR instead of AIR REG.
The compression ability of Parquet varies for each column, and the workers only read in the columns relevant to the query. To determine the \textbf{tuple data size} of a read, we determine the mean entry size of each column by inspecting the Parquet files. For each table, we inspect the first partition at SF=100, which yields us the total byte size of each column and the total number of records. We obtain the size of a column by simply dividing the two. Using this method, we determined column \texttt{l\textunderscore extendedprice} of LINEITEM (size: SF $\times$ 6M tuples) has on average 5.26 byte per tuple. This is of use during estimation: for example in a scenario of SF=200 with 384 partitions this means retrieving the column from a single partition will result in $200\times 6\text{M}\times 5.26 / 384 = 15.7$~MiB the model expects to be retrieved. We use the 1M-reduction \textbf{process rate} from \S\ref{sec:applied-model} for all except Q6, which is the only query of the six which does not have a join or reduce-by-key operation. For Q6, we set the model process rate to a value higher than any possible network rate, which means it is of no consequence as the minimum of the two rates is taken to calculate input duration in the model (\S\ref{sec:model-input}).


\begin{figure*}[t]
	\centering
	\includegraphics[width=\textwidth]{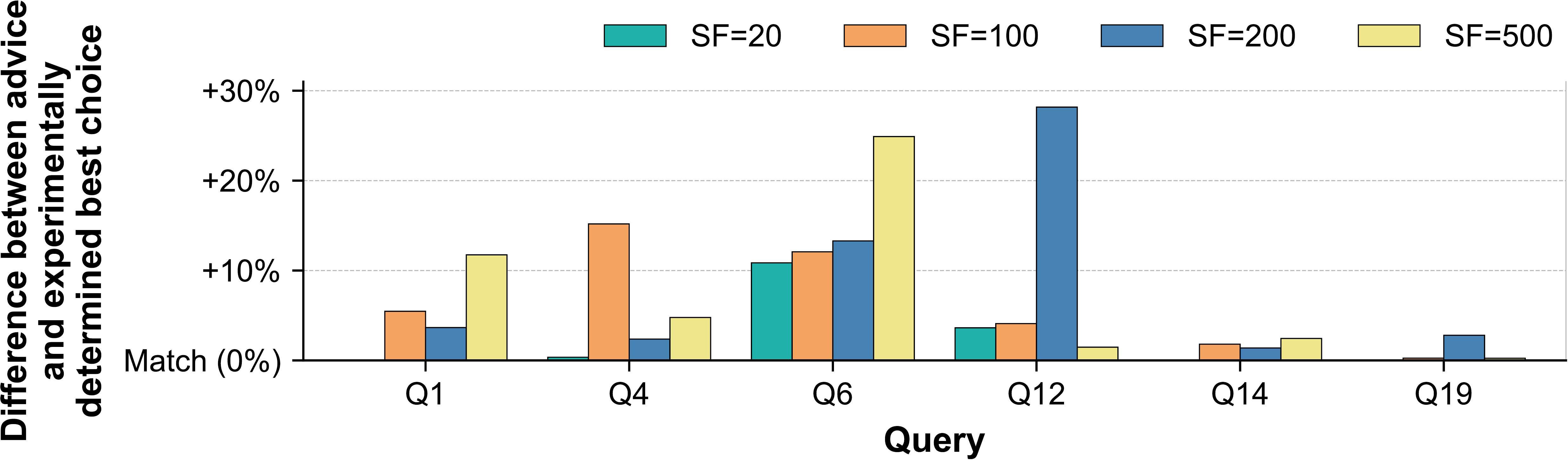}
	\vspace{-14pt}%
	\caption{Advice outcome performance. The bar height is the difference in the distance metric value (which combines completion time and cost) between the advised configuration and the experimentally determined best configuration.}
	\label{fig:tpc-h-bars}
\end{figure*}

\subsection{Experimental setup}

We evaluate across four scale factors $SF\in\{20, 100, 200, 500\}$ (see Tab.~\ref{tab:tpc-h-size}) for six queries (Q1, Q4, Q6, Q12, Q14, Q19). For each $(SF, Q)$ combination, we perform runs across 49 configurations: the Cartesian product of seven number of workers and seven memory sizes (768, 1024, 1280, 1769, 2048, 2560, 4096). The seven numbers of workers are determined via the targeted number of input partitions ($F$) per worker: $F\in\{1, 2, 3, 4, 8, 16, 32\}$ For example, at $SF=100$ with 192 partitions, the number of workers is $W\in\{\frac{192}{32}, \frac{192}{16}, \frac{192}{8}, \frac{192}{4}, \frac{192}{3}, \frac{192}{2}, \frac{192}{1}\}=\{6, 12, 24, 48, 64, 96, 192\}$. There is one exception for $SF=500$, at which we do not run $F=1$ (W=960) as it is costly and clearly not in the regime of interest. Any configuration for which any of its repeated runs results in out-of-memory or reached time limit are considered infeasible (26 out of 1134, thus we have 1108 total configurations). Each configuration is run three times, and the best data point is chosen in accordance with \S\ref{sec:evaluation-practice}.


\subsection{Results}

\begin{figure*}[t]
	\begin{subfigure}[t]{0.48\textwidth}
        \centering
		\includegraphics[width=\textwidth]{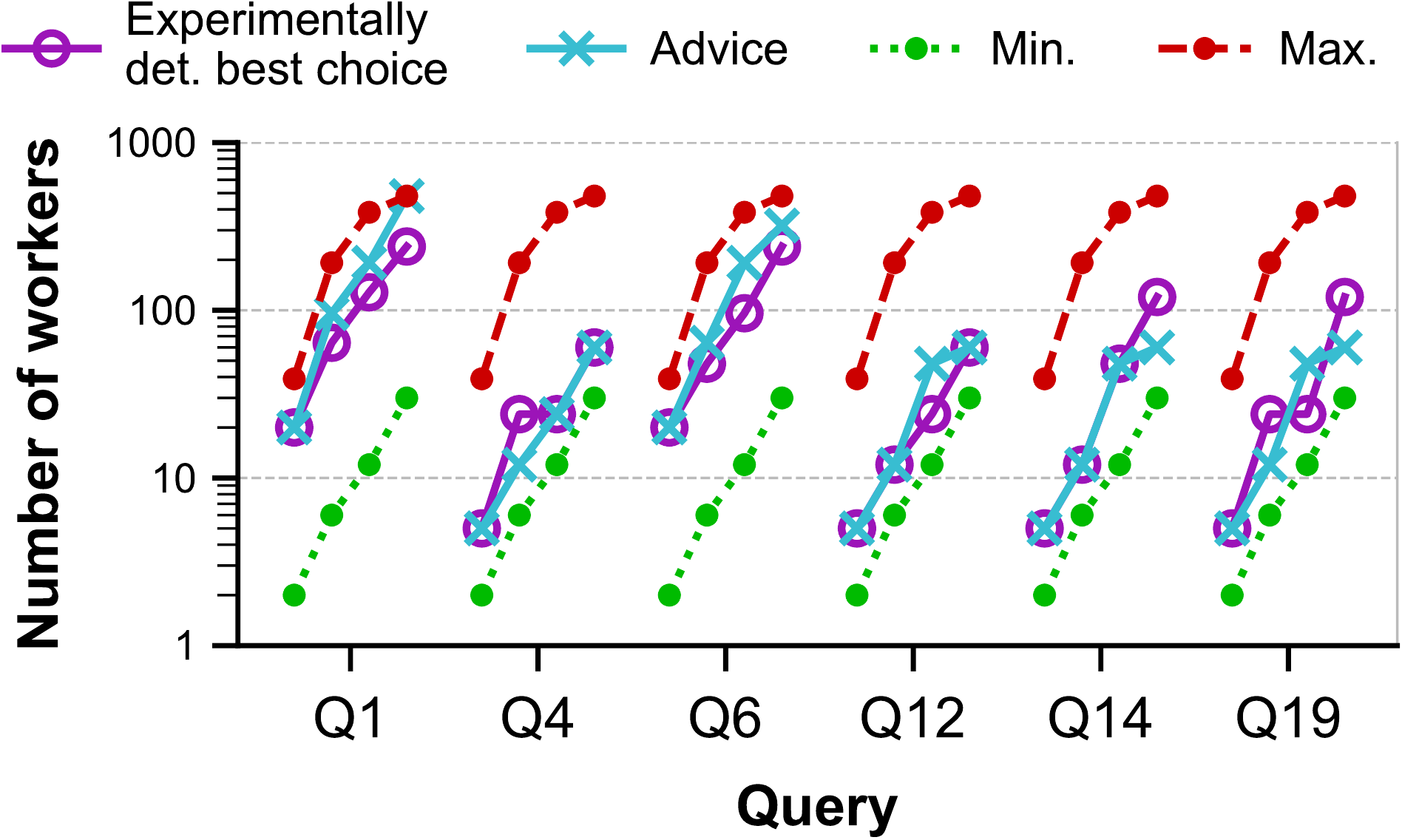}
		\caption{Number of workers (log-scale).}
		\label{fig:tpc-h-choices-num-workers}
    \end{subfigure}%
 	\hfill%
	\begin{subfigure}[t]{0.48\textwidth}
        \centering
		\includegraphics[width=\textwidth]{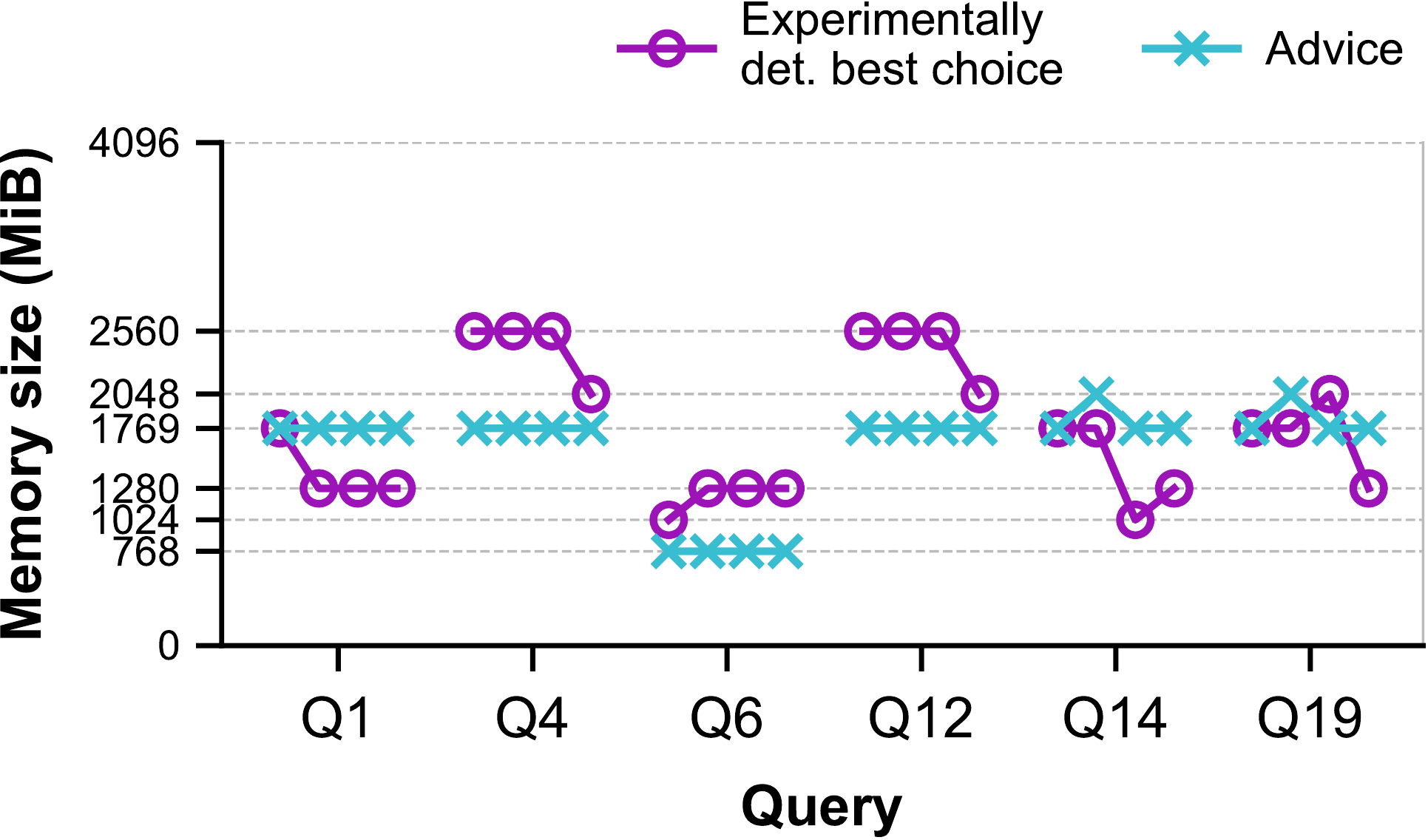}
		\caption{Memory size.}
		\label{fig:tpc-h-choices-memory-size}
    \end{subfigure}%
	\vspace{-2pt}%
	\caption{The choices made by the advisor and from the experiments. For each query, the four scale factors are plotted left to right.}
	\label{fig:tpc-h-choices}
	\vspace{-6pt}
\end{figure*}

The main aspect to explore is how close the recommended configuration given by the advisor is to the best configuration determined experimentally. For this, we use the metric as defined in the evaluation methodology (\S\ref{sec:advice-evaluation}): this distance metric combines the achieved completion time and financial cost, and produces a single distance value. We normalize the distance by that of the experimentally chosen best configuration distance (\ie the best observed run). We show this value for all settings in Fig.~\ref{fig:tpc-h-bars}. The configurations determined best experimentally and the advisor choices are shown in Fig.~\ref{fig:tpc-h-choices}. The way to read this information is as follows:\\

\textit{For Q14 at SF=100, the advice is the same number of workers (Fig.~\ref{fig:tpc-h-choices-num-workers}) but with more memory (Fig.~\ref{fig:tpc-h-choices-memory-size}) than the actual best configuration. This results in an outcome 2\% worse overall (Fig.~\ref{fig:tpc-h-bars}) as it finishes 2\% faster but at 9\% higher cost.}\\

Although the suggested configuration not always matches the best experimental run, the advisor chooses a configuration that is within 15\% off the best one determined experimentally 21 out of 24 times. This is sufficiently accurate in practice given the large amount of uncertainty inherent to cloud execution and the problem at hand (see below for the noise inherent to the experimental runs). The larger differences observed for some queries and scale factors have two main reasons.

\begin{figure}
 \begin{center}
  \includegraphics[width=0.5\textwidth]{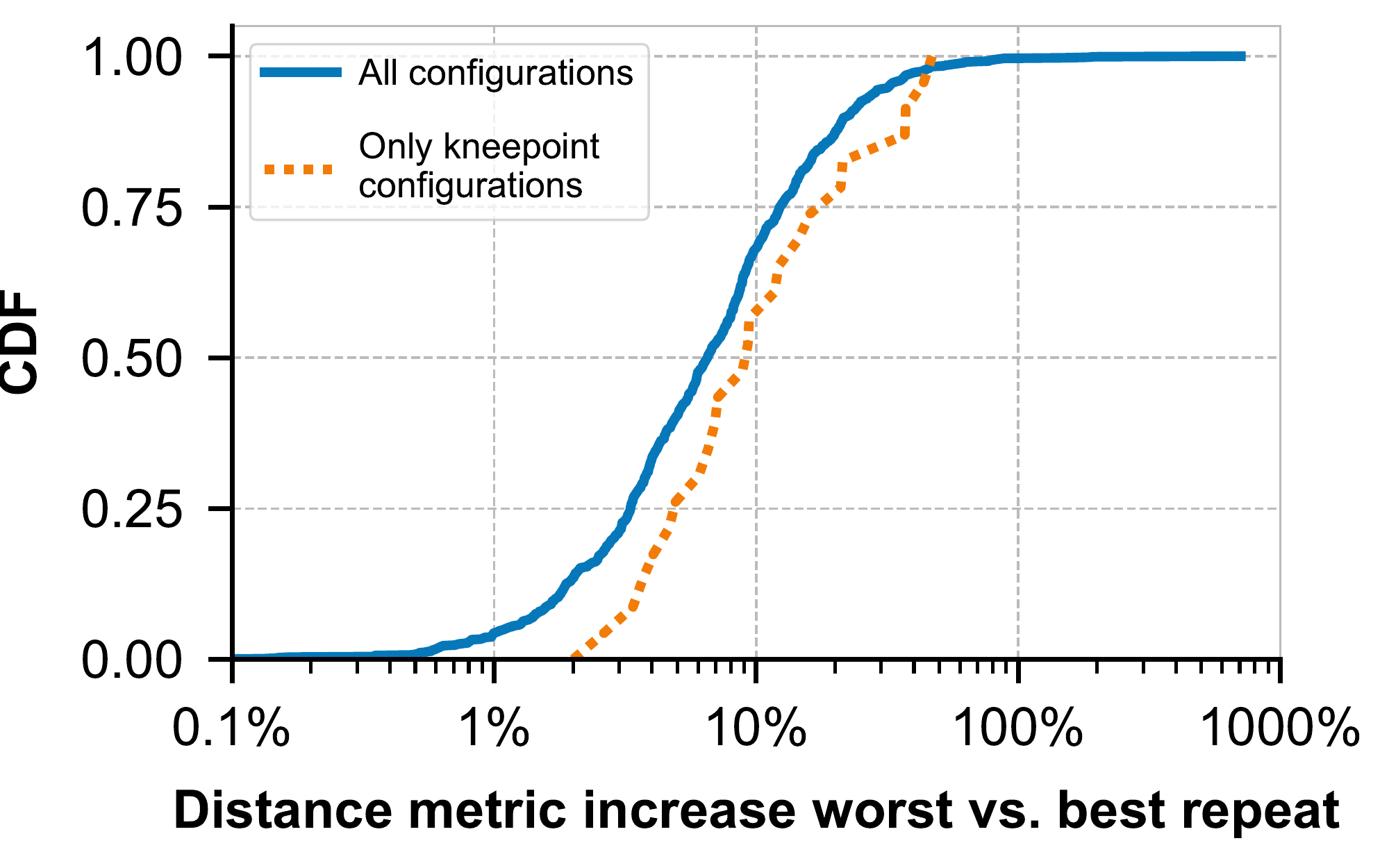}
  \caption{CDF comparing the performance of the worst and best repeated run of each configuration.}
  \vspace{-6pt}
  \label{fig:tpc-h-actual-variance}
  \vspace{-6pt}
 \end{center}
\end{figure}

The first reason is the inherent service variance of the cloud. Especially the start-up time of workers and network rates when reading files exhibit large fluctuations and can occasionally experience very low performance. This is influential because of (a) the completion time being in the order of seconds where even a small delay immediately causes a large delay in percentage, and (b) the dependency on the tail: the completion time is determined by the last worker finishing and different forms of stragglers occur often, especially at larger scales. To illustrate this effect, we compare in Fig.~\ref{fig:tpc-h-actual-variance} the distance metric for the best and the worst of the three runs of each of the 1108 configurations (contiguous line) and the subset of kneepoints (dotted line). For the kneepoints (the points we use as reference): half have a variance of more than 10\% between best and worst and almost a quarter of them have a difference of over 20\%. This is why we consider that an advice that is within 15\% of the best observed run is a more than reasonable approximation of a suitable configuration. In current systems, it is impossible to get more accuracy but the advice is still valuable in automating the process of deploying queries on serverless platforms. 

The model limitations and assumptions are the second reason for differences between estimation and actual runs. The goal of the model is to capture the relative performance impact of the various components (\ie start-up, input, exchanges, ...). Certain aspects of the Lambada and AWS are difficult to estimate and are often not needed to capture the overall trend. Yet, in some cases, these can have an impact in advice outcome. In particular, issues such as the temporary bursty higher network rate when having more than one vCPU, and the effect of concurrent processing and networking on their respective rates cause unusual experimental results that are infeasible to model. 

Although the first reason above is always present, the latter reason is of most interest to explain the two cases in Fig.~\ref{fig:tpc-h-bars} which are more than 20\% off.

In the first case, Q6 at SF=500 has a low completion time, with the best (kneepoint) choice finishing in 2.7~s. This amplifies any delay in terms of relative outcome: the advice outcome is 3.9~s (+43\%) which is far from the best run but close to the other runs. Secondly, the model overestimates network and process performance of low memory workers, and as a result prefers the cheaper lower memory size. The preference for higher number of workers stems from modeling the worker start-up as a constant (it actually grows with the number of workers but it is difficult to predict by how much, introducing variances in the order of hundreds of milliseconds), and with increased probability of encountering stragglers at higher worker numbers (with stragglers being completely unpredictable). The second case, Q12 at SF=200, has an advice whose outcome is 28\% off the best run with a completion time 12\% slower and 41\% more expensive. The model underestimates the increased processing and network rate at higher memory size, and is too optimistic regarding the additional overhead changing from a 1-level to a 2-level exchange (with 48 instead of 24 workers exceeding the threshold of 32), which experimentally does not yield a sufficiently cost-effective decrease in completion time.

Some of these aspects can be corrected an accounted for by modifying the constants or specializing the model even more for exchange heavy or scan heavy queries. We did not think it was effective to do so given the level of noise that is inherent to the system. In addition, if very many repetitions of the same query are run, a more realistic picture of its actual performance emerges but the number of runs is very high and the variance remains very high. For a system like Lambada, which targets occasional queries over cold data, the average over many runs is not relevant because queries will be run only once or twice and the probability that they run into the situations described is high as our own experiments with three runs for each configuration indicate (Figure ~\ref{fig:tpc-h-actual-variance}). 

\greybox{
Across the more representative workloads, we observe the following:
\begin{itemize}[leftmargin=10pt,itemsep=2pt,topsep=2pt]

    \item There is significant variance in the outcome of configurations as serverless data processing systems depend on the cloud for their start-up and communication performance, yet it is still possible to give sound advice;
    
    \item The model is able to suggest a configuration with a good balance of completion time and financial cost: in the vast majority of cases it was less than 15\% off the best choice a margin within the noise of the system.
    
\end{itemize}
}

\section{Related work}
\label{sec:related-work}

There is a lot of work on resource allocation in the cloud for VMs \cite{Microsoft-Protean20} and containers \cite{Pop-allocation21}. Recent work from Microsoft has explored how to map serverless to the underlying VMs running functions from the perspective of the cloud provider \cite{serverless-to-VM21} but that is a different problem from the one we address in this paper. There is also work in trying to explore the properties of serverless systems \cite{Microsoft-study20,Peeking-ATC18}. Existing serverless data processing systems~\cite{lambada, starling, boxer} show how to outperform alternatives in both completion and cost in select configurations, but use manually selected configurations.

Existing work on rough estimation of execution time and cost for VMs~\cite{leis2021towardscostoptimal} does not directly translate to serverless functions due to: (a) the difference in the number of workers involved (10s for virtual machines, 100s for serverless functions), (b) the overhead incurred for data exchanges (typically through storage), and (c) the difference in task duration. Serverless operates in the few to tens of seconds range, whereas virtual machines in minutes to hours -- factors insignificant at a one hour scale become so at a one second scale \cite{Peeking-ATC18,Microsoft-study20}. Similar to \cite{leis2021towardscostoptimal}, we account for processing (albeit using data and rate instead of CPU hours) and network overhead, but also incorporate the factors unique to serverless query processing systems, namely startup time, exchange overhead and request cost. Another related work is Astr(e)a~\cite{astra, astrea}, which similarly proposes a parameterized model to estimate and advise serverless configurations. However, their model focuses on long running analytic queries and thus does not incorporate startup time and exchange overhead. Their configuration advice minimizes completion time or cost provided a user-defined budget or performance constraint respectively, rather than a mechanism as ours which determines a tradeoff between completion time and cost.

In the context of Apache Spark, Sparklens~\cite{sparklens} is a tool which estimates completion time and utilization given a certain number of compute nodes after inspecting a single run -- especially useful for repeated queries but not applicable in our use case. In the context of Azure Synapse, \cite{sen2021predictivepriceperf} proposes the AutoExecutor framework, built upon SCOPE~\cite{scope}, to automatically choose the number of executors for its Spark SQL queries. TASQ~\cite{anish2021optimalresource} which similarly is built on top of SCOPE~\cite{scope}, trains and makes use of (graph) neural networks to optimize the amount of tokens allocated to a job. Song et al.~\cite{song2020boosting} proposes a multi-objective optimization method to tune several Spark parameters including number of executors and cores, as well as memory allocation. In the work, they ran a variety of run configurations and used its resulting traces to train for each workload neural network models for various objectives including latency and cost. Besides the different target platform, these works differ from ours in the design of the model. Our goal is to provide an advisory tool with a rough estimation model based on reasoning with parameters that can be adjusted for any serverless query processing system or particular query, rather than applying learning from traces with performance metrics to determine the best allocation. Our model incorporates the most important dimensions for serverless query processing in the cloud, namely start-up, compute, network and exchange overhead. \cite{song2020boosting} similarly made use of Pareto set to determine the best configuration, although the method at which the Pareto frontier is formed (via multi-objective optimization) differs from ours (which simply estimates the entire Cartesian product of possibilities), as well as the strategy of choosing the best configuration differs as they use direct distance to the idealized fastest and cheapest point rather than using that point as normalization for distance to the origin as we do.


\section{Conclusion}
\label{sec:conclusion}

We have presented an advisory tool for serverless data processing designed to pick a configuration (\ie number of workers and their memory size) striking a good balance between performance (completion time) and cost. The advisory tool makes use of a rough estimation model which incorporates processing and networking, start-up, exchange overhead and request cost. We evaluated the advisor's performance using an existing serverless data processing system and show it is able to identify desirable configurations. Automated configuration facilitates the job of the user and broadens the scope at which serverless data processing is cost-effective.

\balance


\bibliographystyle{ACM-Reference-Format}
\bibliography{999-bibliography}

\end{document}